\documentclass[12pt,preprint]{aastex}	

\begin{document}                           			
\title{Chandra Observation of the Cluster Environment of a WAT Radio Source in Abell 1446}  							
\author{E.M. Douglass\altaffilmark{1},
            Elizabeth L. Blanton\altaffilmark{1},
            T.E. Clarke\altaffilmark{2,3}
            Craig L. Sarazin\altaffilmark{4},
		Michael Wise\altaffilmark{5}}

 	\altaffiltext{1}{Institute for Astrophysical Research, Boston University, Boston, MA 02215}
    	\altaffiltext{2}{Naval Research Laboratory, Washington, DC. 20375}
    	\altaffiltext{3}{Interferometrics, Inc., 13454 Sunrise Valley Dr. No. 240, Herndon, VA 20171}
	\altaffiltext{4}{Department of Astronomy, University of Virginia, P.O. 400325, Charlottesville, VA 22904-4325, USA} 
	\altaffiltext{5}{Astronomical Institute, Anton Pannekoek, NL-1098 SJ Amsterdam}

\begin{abstract}
Wide-angle tail (WAT) radio sources are often found in the centers of galaxy clusters where intracluster medium (ICM) ram pressure may bend the lobes into their characteristic C-shape.  We examine the low redshift ($z=0.1035$) cluster Abell 1446, host to the WAT radio source 1159+583.  The cluster exhibits possible evidence for a small-scale cluster-subcluster merger as a cause of the WAT radio source morphology.  This evidence includes the presence of temperature and pressure substructure along the line that bisects the WAT as well as a possible wake of stripped interstellar material or a disrupted cool core to the southeast of the host galaxy.  A filament to the north may represent cool, infalling gas that's contributing to the WAT bending while spectroscopically determined redshifts of member galaxies may indicate some component of a merger occurring along the line-of-sight.  The WAT model of high flow velocity and low lobe density is examined as another scenario for the bending of 1159+583.  It has been argued that such a model would allow the ram pressure due to the galaxy's slow motion through the ICM to shape the WAT source. A temperature profile shows that the cluster is isothermal (kT= 4.0 keV) in a series of annuli reaching a radius of $\sim$ 400 kpc. There is no evidence of an ongoing cooling flow.  Temperature, abundance, pressure, density, and mass profiles, as well as two-dimensional maps of temperature and pressure are presented.  
\end{abstract}


\section{INTRODUCTION}  

Named for their characteristic C-shape, WATs are double-lobed sources of intermediate radio luminosity typically associated with the central dominant cD or D type galaxy within a galaxy cluster.  It had been originally assumed that the WAT shape was due to the host galaxy's swift motion through the ICM \citep{OR76}. This is seen with Narrow Angle Tail (NAT) galaxies as they plow through the gas in the outer regions of clusters \citep{Miley72} with large peculiar velocities.  The canonical view of WATs puts them at or near the center of their cluster's potential well with low peculiar velocities ($v$ $\leq$ 300 km s$^{-1}$) relative to the intracluster medium.  Given the low peculiar velocities and the assumed densities and velocities of the particles within the WAT sources \citep{Eilek84, ODon93}, this NAT analogy seems to fail as a scenario to produce the bent tail morphology in WATs.

It has been argued that cluster mergers can produce the ICM ram pressure necessary to shape WAT sources \citep{Burns94, Pinkney95, Gomez97, SM00}.  Elongations of a cluster's X-ray emission along the line that bisects the WAT \citep{Gomez97}, significant offset ($\geq$ 100 kpc) of the WAT from the X-ray centroid \citep{SM00}, and the presence of X-ray sub-structure \citep{Burns94, Gomez97} have all been observed in WAT-hosting clusters and are assumed to be indicators of a recent cluster merger.  Simulations have shown that relative velocities of the ICM, great enough to bend WATs, are produced in mergers and persist on time scales greater than the lifetime of the radio sources (10$^7$-10$^8$ years) \citep{Roettiger93, Loken95, Roettiger96, RS01}. 

Large velocities relative to the ICM resulting from cluster mergers may not be required to bend the WAT if a model of high flow velocity and low density is assumed within the radio lobes \citep{BB82, Eilek84, ODon90, Pinkney95, Hardcastle05}.  \citet{Hardcastle05} assume this model arguing that galaxy velocities as low as 100-300 km s$^{-1}$ are sufficient to produce the WAT morphology in some clusters. In previous models as mentioned above, denser lobes with lower flow velocities require galaxy velocities greater than 1000 km s$^{-1}$.  

Abell 1446 is a low redshift ($z=0.1035$) galaxy cluster of richness class 2.  The bright central galaxy is a large elliptical ($M_{\rm v}=-23.4$) which is host to the radio source 1159+583, one of the six objects originally defining the WAT class in \citet{OR76}.  Abell 1446 has been previously imaged by both the \emph{Einstein} \citep{BB82} and \emph{ROSAT} observatories \citep{Gomez97}.  Here, we report on the results of a 59 ksec \emph{Chandra} observation of Abell 1446, which is the highest resolution observation, spatially and spectrally, of the cluster to date.  The greatly improved spatial resolution is necessary to probe the substructure of the ICM within the central 100 kpc where the radio lobe and ICM interaction is taking place. The WAT production mechanisms discussed above will be investigated for this cluster.  Unless otherwise noted, all uncertainties are given to 90\% accuracy.  We assume $\Omega _{\Lambda} =0.7$, $\Omega _{M}=0.3$, and $H_{\rm o}=70$ km s$^{-1}$ Mpc$^{-1}$.  At $z=0.1035$, the angular diameter distance is $D_{\Theta}=392.14$ Mpc, the luminosity distance is $D_{\rm L}=477.53$ Mpc, and $1\arcsec = 1.9$ kpc.

\section{OBSERVATION AND DATA REDUCTION}
\subsection{\emph{Chandra} Observation}
Abell 1446 was observed by \emph{Chandra} on 2004 September 29 for a total of 59,120 seconds in very faint (VFaint) mode.  The center of the cluster was positioned on the S3 chip of \emph{Chandra}'s ACIS-S detector with a 1$\arcmin$  offset from the nominal pointing to avoid node boundaries. The data analysis package, CIAO version 3.2, was used for data reduction.  By analyzing light curves of the S1 and S3 chips throughout the observation period, it was determined that no background flares occurred to contaminate the data. The data were corrected for hot pixels and cosmic ray afterglows using standard techniques. Background files were taken from the blank sky observations of M. Markevitch included in the CIAO calibration database (CALDB) and reprojected to match the A1446 observation.  

\subsection{Radio Reductions}
Radio data presented in this paper for 1159+583 were extracted from the NRAO VLA archive. Observations in A configuration at 1452.4 and 1502.4 MHz were taken in 1986 March 26 (VLA code AO62), while C configuration observations at 1464.9 and 1514.9 MHz for the same project code were taken in 1985 July 17. A second frequency band was also reduced in order to study the spectral properties of the radio source. The data were taken for CLA code AO49 at 4835.1 and 4885.1 MHz in 1984 February 17 in the VLA B configuration and 1984 August 5 in the VLA D configuration.

All observations used 3C286 as a flux calibrator and data were calibrated and reduced with the NRAO Astronomical Image Processing System (AIPS). Images at each configuration were produced through the standard Fourier transform deconvolution method. All data sets were processed through several loops of imaging and self-calibration to reduce the effects of phase and amplitude errors in the data. The A and C configuration observations were added together to produce a final image with an rms of 82 $\mu$Jy/bm while the combined B and D configuration data has an rms of 31 $\mu$Jy/bm. 

\section{SPATIAL DISTRIBUTION OF CLUSTER EMISSION}
An adaptively smoothed image of the cluster 800 kpc (421$\arcsec$) on a side is shown in Fig. 1.  It was created using the \emph{csmooth} task in CIAO with a minimum SNR of 3.  It has been corrected for background and exposure. The photon energy range has been restricted to 0.3-7.0 keV. From this image it can be seen that some substructure is present.  While on larger scales the cluster appears to be approximately circularly symmetric, the inner 200 kpc (105$\arcsec$) show anisotropies in the distribution of the gas.  Most prominently, there appears to be a filamentary excess of emission to the north of the AGN and a bright area of emission to the southeast.  

The contours of the smoothed cluster image have been overlaid on the Sloan Digital Sky Survey image (Fig 2.).  The peak of the X-ray emission corresponds with the large elliptical galaxy which is the host of the WAT and approximately at the center of the cluster.   

\subsection{Point Sources}
Individual sources were detected using the wavelet detection algorithm \emph{wavdetect} in CIAO. Of the 20 sources that were detected, four correspond with sources in the USNO A2.0 catalog.  Those sources that were identified in the \emph{Chandra} image are the central $R=15.69$ mag elliptical galaxy, an $R=17.7$ mag galaxy 100$\arcsec$ to the northeast of the central galaxy, an $R=19.9$ mag galaxy 125$\arcsec$ to the west-northwest, and an $R=16.9$ mag galaxy 125$\arcsec$ to the northwest.  All are cluster members.  The four \emph{wavdetect} sources were identified within 1$\arcsec$ of their optical counterparts suggesting that there is no positional offset of the X-ray image.  Therefore no astrometric correction has been applied to the data.  These sources can be seen as the red squares in the Sloan Digital Sky Survey (SDSS) image of the cluster (Fig. 2).  

\subsection{Cluster Center and Ellipticity}
To determine the center and projected ellipticity of the X-ray emission, isophotes were fit to a 2$\arcsec$ Gaussian smoothed image of the cluster.  This was done using the \emph{ellipse} task, included in the STSDAS package of IRAF. The center position, eccentricity, and pitch angle were all allowed to vary. The point sources were replaced with values interpolated from the surrounding regions using \emph{dmfilth} in CIAO.  The \emph{ellipse} task found two center positions, depending on the radius of the isophote. The centers of the isophotes with semi-major axes $\leq$ 45$\arcsec$ are coincident with the AGN (X-ray point source coincident with the radio core of 1159+583) while those isophotes with greater semi-major axes are centered near a position 25$\arcsec$ to the northeast.  As a test, the analysis techniques discussed in this paper were performed assuming the cluster center is coincident with the AGN, the position 25$\arcsec$ to the northeast of the AGN, and points in-between.  No significant variation in the result was observed using the different cluster centers.  With the majority of the bright central emission centered about the WAT, except where noted, the AGN is used as the center of the cluster throughout the paper.  The position 25$\arcsec$ to the northeast of the AGN is considered to be the center of the more diffuse large-scale emission.  Due to the low ellipticity ($e \leq 0.16$) of the isophotes with radii larger than 45$\arcsec$, it was decided that A1446 may be treated as roughly circularly symmetric.  This assumption allows for the use of circular annuli in the analysis process discussed below.

\subsection{Surface Brightness Profile}
Assuming large-scale circular symmetry, a surface brightness profile in the energy band 0.3-10.0 keV was extracted from 22 concentric annuli centered on the AGN with each annulus containing at least 1000 source counts (Fig. 3). The profile was corrected for background and exposure with the point sources, including the AGN, excluded.  A one-dimensional beta model was fitted to the data in CIAO's modeling and fitting application \emph{Sherpa}. The fit is good with a reduced $\chi^2$ = 0.51 with 19 degrees of freedom.  The model parameters are $\beta = 0.43 ^{+0.075}_{-0.041}$ with a core radius $R_c = 79.76^{+29.67}_{-16.51}$ kpc and a peak amplitude of 0.089$^{+0.016}_{-0.006}$ cts s$^{-1}$ arcmin$^{-2}$.  While the beta values overlap within the errors, the core radius differs significantly from that determined in \citet{Gomez97} using \emph{ROSAT} data where $\beta$ = 0.7$\pm$0.2 and $R_{c}=260\pm14$ kpc.  This discrepancy is most likely due to \emph{ROSAT}'s larger field of view.  Surface brightness profiles of clusters of galaxies tend to be best fit by a double $\beta$-model and rich clusters such as A1446 tend to exhibit a steeper surface brightness profile at large radii ($\beta \sim 0.7$).  Here we are fitting the surface brightness profile of the cluster core as opposed to the large-scale cluster emission as had been done by \citet{Gomez97}.  A double beta model was fit to the data.  The parameters for the two models were calculated to be $R_{c,1}=230^{+1300}_{-100}$ kpc, $\beta_1 = 0.7^{+2.7}_{-0.21}$ and $R_{c,2}=290^{+62.7}_{-209}$ kpc, $\beta_2 = 7.3^{+2.7}_{-3.9}$ with the peak amplitudes being 0.05 and 0.04 respectively. For this fit the reduced $\chi^2$ = 0.32 with 16 degrees of freedom.  Though the values of the model match those of \citet{Gomez97} more closely, the uncertainties are very large.

\subsection{ICM and WAT Interaction}
The central elliptical galaxy within Abell 1446 is host to the prototypical WAT radio source 1159+583.  The source consists of two radio lobes bending in the characteristic C-shape that is typical of WAT sources.  The center of the radio emission is coincident with the area of brightest X-ray emission in the cluster and within 25$\arcsec$ (47 kpc) of the center of the large-scale cluster emission determined above.  This suggests that the host galaxy, as the largest member of A1446, has sunk nearly to the bottom of the cluster's gravitational potential well.  While the southern lobe's projected length extends to a distance of roughly 105 kpc (55$\arcsec$) from the host galaxy, the northern lobe ends only 45 kpc (24$\arcsec$) northeast of its host in projection.  Both the northern and southern lobes seem to be bending towards the southeast. On close visual inspection (Fig. 4), the northern lobe appears to have carved out a cavity within the cluster gas.  An excess of emission is seen between the lobes.     
    
\subsection{Cluster Substructure}
As can be seen in the smoothed image of the cluster (Fig. 1), the X-ray emission in the central regions is not symmetrically distributed about the cluster center.  This was investigated more closely by fitting the smoothed X-ray image with an isophotal model.  A Gaussian smoothed ($\sigma$=4$\arcsec$) image was used as input in the IRAF $\emph{ellipse}$ task.  The ellipticity and position angle were free to vary while the center was fixed at an intermediate value between the AGN and the large-scale center. When fitting the central 150 kpc (79$\arcsec$) the position angle varied to some degree suggesting the presence of substructure in the X-ray gas as had been previously seen in \citet{Gomez97}.  Using \emph{ROSAT} data, \citet{Gomez97} detected the excess of emission to the northeast of the cluster as a significant X-ray clump. An excess to the west of the center was also detected.  A smooth elliptical model of surface brightness was created from the table of elliptical isophotes returned by the \emph{ellipse} task.  This model was then subtracted from the Gaussian smoothed image to reveal any excess emission.  The model was weighted by 0.7 in order to avoid over subtracting.  This residual image is shown in Fig. 5.  An excess can be seen directly north of the AGN as well as to the southeast.  The radio lobes correspond to regions of reduced emission (particularly the northern lobe).

\section{SPECTRAL ANALYSIS}
\subsection{Total Spectrum}
A spectrum was extracted from a region of radius 190$\arcsec$ around the center of the cluster using ACISSPEC in the CIAO package.  The central AGN along with other X-ray point sources mentioned above were excluded.  The spectrum was binned such that each energy bin contained a minimum of 40 counts after the background had been subtracted.  Restricting the energy range to 0.7-7.0 keV, a number of models were fitted to the spectrum using XSPEC V11.3.2.  The three models that were fitted to the spectrum were: a single temperature model accounting for the thermal bremsstrahlung emission of the hot ICM including line emission from several elements (1-MEKAL), a MEKAL model accounting for two populations of hot gas at different temperatures (2-MEKAL), and a cooling flow model (MKCFLOW) combined with a single temperature MEKAL model.  Temperature, abundance, and normalization were free parameters. Each model was run twice, with the absorption parameter free to vary and then fixed at the Galactic value of 1.5 $\times$ 10$^{20}$ cm$^{-2}$ \citep{DL90}.  The lack of an improved fit with the addition of a second MEKAL component or an MKCFLOW component is consistent with the cluster ICM being comprised of nearly isothermal gas.  The average temperature of the inner 360 kpc of the ICM was determined to be 4.00 $\pm$ 0.19 keV with an average abundance value of 0.34 $\pm$ 0.07 Z$_{\odot}$ from the single MEKAL fit with fixed Galactic absorption. 

\begin{table}[h]
\begin{center}
{Table 1: XSPEC fits to the X-ray spectrum of the inner 190$\arcsec$ radius region}\\
{\scriptsize{Note: numbers within parenthesis indicate frozen values.}}\\
{ }
{\scriptsize
\begin{tabular}{l cccccc}
\hline\hline
{}&{$N_{H}$}& {$kT_{\rm low}$}&{$kT_{\rm high}$}&Abundance&$\dot{M}$&\\
{Model} & (10$^{20}$cm$^{-2}$)&{(keV)}&{(keV)}&(Z$_{\odot}$)&M$_{\odot}$yr$^{-1}$&$\chi ^{2}$/dof \\ [0.3ex]
\hline\hline
MEKAL & (1.50) & $\dots$& 4.00$\pm0.19$ &  0.34$\pm0.07$ &$\dots$& 345/308 = 1.12\\
MEKAL  & 2.12$\pm 1.3$ & $\dots$& 3.92$\pm 0.25$ &  0.33$\pm0.07$ &$\dots$& 344/307 = 1.12\\
MEKAL+MEKAL & (1.50) & 1.8$^{+78.1}_{-1.7}$ & 4.26$^{+75.6}_{-0.08}$  &  0.33$\pm0.07$ &$\dots$& 344/306 = 1.13\\
MEKAL+MEKAL& 2.78$^{+1.66}_{-1.51}$ & 1.93$^{+78.0}_{-1.84}$ & 4.48$^{+75.4}_{-4.40}$  &  0.29$\pm0.08$ &$\dots$& 344/306 = 1.13\\
MEKAL+MKCFLOW& (1.50) & 1.80$^{+4.10}_{-1.72}$ & 4.70$^{+3.58}_{-0.84}$  &  0.33$\pm0.07$ &56.78$^{+3391}_{-56.78}$ & 344/306 = 1.13\\
MEKAL+MKCFLOW& 2.71$^{+1.75}_{-1.44}$ & 1.80$^{+4.10}_{-0.51}$ & 5.59$^{+2.82}_{-1.78}$  &  0.30$\pm0.07$ &93.03$^{+1183}_{-93.03}$ & 342/305 = 1.12\\

\hline
\end{tabular}}
\end{center}
\end{table}

The X-ray temperature measurement is in agreement with previously derived temperatures for the A1446 ICM in \citet{BB82}, $kT=4^{+5}_{-2}$ keV, and \citet{Gomez97}, $kT=2.6^{+1.3}_{-0.7}$ keV, using \emph{Einstein} and \emph{ROSAT} data, respectively.  While the \citet{Gomez97} value is slightly lower than the value determined in this paper, they agree to within the errors.  
  
\subsection{Temperature and Abundance Profiles}
Spectra were extracted from 10 concentric annuli centered on the AGN.  Each annulus had roughly 2500 net source counts with the central AGN and other X-ray point sources excluded.  As can be seen in Fig. 6, the temperature remains roughly constant (4.0 keV) out to a radius of 400 kpc (210$\arcsec$). This is an indication that the cluster is not the site of a cooling flow.  Cooling flow clusters typically exhibit a sharp decrease in temperature within the central region to roughly one-half the value in the outer regions \citep{Peterson03}.  There is also an absence of any significant temperature discontinuities that would result from a recent large-scale cluster-cluster merger.  There is a slight jump in temperature at a radius of 150 kpc (79$\arcsec$) which may correspond to a hot subclump of gas present in the ICM.  

An abundance profile was extracted using five concentric annuli centered on the AGN.  As can be seen in Fig. 6, the profile values and uncertainties restrict the abundance to the range 0.20-0.55 Z$_\odot$ in the four inner annuli.  The radial profile for the chemical abundance does not show any definite radial dependence with an average value of $\sim$ 0.34 Z$_\odot$.

The values of temperature and chemical abundance determined from the radial profiles agree well with the average values determined from the best-fitting single-temperature MEKAL model of the total cluster spectrum.  Due to the uncertainties in the abundances, the average value determined from the total cluster spectrum was used below ($\S 4.3$).  

\subsection{Temperature Map}
A temperature map was created of a region 375 kpc (197$\arcsec$) on a side centered on the AGN.  Each value on the temperature map was determined by extracting a spectrum from the smallest surrounding region on the chip that contains 1000 total counts.  Each spectrum was fit to a single temperature MEKAL model with the value of the chemical abundance frozen at 0.34 Z$_{\odot}$, the value from the fit to the whole cluster.  The two dimensional map was then smoothed with a Gaussian of $\sigma$=4$\arcsec$.  The temperature map can be seen in Fig. 7.  Again, there is no evidence for a cooling flow.  The temperature map agrees well with the temperature profile from above as there is no strong evidence of a radial dependence on temperature.  As seen in the temperature profile, values of temperature are constant with radius when averaged over annuli.  However, the temperature map reveals temperature variations within the cluster.  Such a variation can be seen 150 kpc (79$\arcsec$) to the southeast of the cluster center.  To investigate this further a spectrum was extracted from an elliptical region coincident with the region of enhanced temperature seen in the map.  The temperature of this region, with semimajor and minor axes of 25$\arcsec$ and 12$\arcsec$ and position angle 20$^\circ$ west of north, was determined to be $kT=5.93^{+2.06}_{-1.21}$ keV.  The temperature of the surrounding gas in a 60$^\circ$ wedge of radius 197$\arcsec$ was found to be $kT=3.72^{+0.47}_{-0.33}$.  This region lies along the line bisecting the bent radio source and will be discussed in $\S 6.1.1$.  Along this same line, but to the northwest of the WAT, there appears to be another region of increased temperature.  Upon further analysis, extracted spectra reveal that the temperature difference is not significantly greater than the neighboring regions.  The uncertainty of the temperature values in the map is roughly 15\% at the center, reaching 40\% towards the edges.
 
\subsection{$L_{X}-T$ Relation}
It has been observed that galaxy clusters exhibit a correlation between their X-ray luminosity and temperature.  Following the method described in \citet{Wu99}, we determined the total bolometric luminosity and global temperature of Abell 1446.  The \emph{Chandra} observation of A1446 only extends out to $\sim$360 kpc (190$\arcsec$) from the cluster center.  The bolometric luminosity of this region was calculated from the total cluster spectrum discussed above accounting for redshift and cosmology.  A `dummy' response file was created to allow the spectral model to be evaluated from 0.001-100 keV.  The contribution of the ICM emission not included on the S3 chip was then estimated.  This was done by extrapolating the \citet{Gomez97} $\beta$-model fit to the \emph{ROSAT} surface brightness out to infinity.  The \citet{Gomez97} model was used as it is more consistent with $\beta$-model fits of rich clusters of galaxies and covers a larger radial extent, requiring less extrapolation, than the \emph{Chandra} profile.  From this analysis it appears that A1446, given its global temperature, has a luminosity falling along the $L_{X}-T$ trend when compared with the \citet{Wu99} sample.  This can be seen in Fig. 8.  From the simulations of \citet{RT02}, clusters undergoing major mergers have boosts in their X-ray luminosity and temperature such that they follow the slope of the relation, but generally fall above the trend line throughout the duration of the merger.  As seen in Fig. 8, Abell 1446 falls slightly below.

\section{Pressure, Density, and Mass}
\subsection{Pressure and Density Profiles}
The pressure and density profiles were calculated using the projected temperatures determined from the radial profile.  Assuming spherically symmetric shells, a routine that deconvolves the X-ray surface brightness was used to give the emissivity as a function of radius.  Assuming the emissivity is due to thermal bremsstrahlung and constant within the spherical shells, the pressure and density were calculated.  As can be seen in Fig. 9, pressure and density generally decrease with increasing radius.  A lack of a sharp jump in the radial profile for pressure indicates an absence of a large-scale shock that would be present if a violent cluster-cluster merger were taking place near the plane of the sky.  While the uncertainties of values along the pressure profile are large enough so that the trend follows a relatively smooth decline with radius, we can establish an upper limit on a possible pressure `discontinuity' at 80".  Moving outward across this point the pressure may decrease by as much as a factor of $\sim$ 1.9, possibly indicating the presence of a weak shock.  From a radius of roughly 400 kpc (210$\arcsec$), the density increases by a factor of 10 to a core density of 3.9 $\times$ 10$^{-3}$ cm$^{-3}$.  The gas pressure also increases by a factor of ten from a radius of 400 kpc (210$\arcsec$) to its core.  The pressure reaches a value of 5.32$^{+0.50}_{-0.49}$ $\times$ 10$^{-11}$ dyne cm$^{-2}$ near the cluster center.  Using \emph{ROSAT} data, \citet{Gomez97} found the pressure near the center to be 7.44 $\times$ 10$^{-12}$ dyne cm$^{-2}$.  This difference may be a result of the larger core radius $\beta$-model from which they determined density, along with their lower global cluster temperature.     

\subsection{Cooling Time}
Using the temperature and density profiles determined above, the cooling time of the intracluster gas can be calculated (Fig. 9).  In relaxed cooling core clusters, the cooling times in the central regions of the cluster are much less than a Hubble time [see \citet{Sarazin88} and references therein].  The short cooling time allows the cooling core to form.  The cooling times of the gas in the central regions of A1446 are greater than the age of the universe, suggesting that there has not been enough time, given the properties of the gas, for a cooling flow to begin. This is in agreement with the flatness of the temperature profile.  The cooling time for the innermost annulus is t$_{\rm cool}$ = 15.5 Gyr.

\subsection{Gas and Gravitational Mass}
The gas and gravitational mass were determined using a routine that deconvolves the X-ray surface brightness of the extended emission to give emissivity and density as a function of radius.  These are integrated to give the gas mass as a function of radius.  The total gravitational mass was derived from the equation of hydrostatic equilibrium.  As can be seen in Fig. 10, the gas mass fraction (ratio of gas mass to gravitational mass, GMF) at the cluster core is  $\sim$0.02 which climbs with increasing radius to $\sim$0.1 at 400 kpc (210$\arcsec$) which is a typical value for the GMF for galaxy clusters at this radius \citep{Allen04}.  The total mass contained within 400 kpc is 5 $\times$ 10$^{14}$ M$_\odot$.  The profile is missing a point at the same radius where there is an increase in pressure.  Since the mass is determined assuming hydrostatic equilibrium, such a change in pressure implies the system is not fully relaxed.  The failure of hydrostatic equilibrium at this radius may be a signature of cluster substructure.   

\subsection{Pressure Map}
A map of projected pressure was created from the temperature map and smoothed image of the cluster.  Assuming the X-ray emissivity, $\epsilon_{X} \propto n_{e}^2$ and $P \propto n_eT$, the square root of the counts image was multiplied by the temperature map \citep{Finoguenov05}.  The values of pressure are arbitrary and the map displays the pressure, in projection, relative to the cluster center (Fig. 11). The pressure map exhibits the radial dependence on pressure seen in the radial profile.  A region of high pressure is seen to the southeast of the cluster center coincident with the temperature enhancement present in the temperature map. There appears to be some correlation between the southern radio lobe and the region of increased pressure with which it is coincident.  The higher pressure region around the cluster center extends towards the northwest and a filamentary region of increased pressure can be seen directly to the north. The anisotropies of the two-dimensional temperature and pressure distributions suggest that subclumps of different temperatures and pressures are present in the ICM.  Such subclumps may be identified as remains of a small-scale merger or infalling or rising cluster gas.     
     
\section{DISCUSSION}
\subsection{Small-scale Merger}
\subsubsection{Merger Along WAT Axis} 
Based on X-ray observations of WAT-hosting clusters it has been argued that large-scale cluster-cluster mergers and smaller scale cluster-subcluster mergers may be responsible for producing the WAT class of radio sources \citep{Burns94, Pinkney95, Gomez97, SM00}.  N-body merger simulations suggest specific observational characteristics are produced as a result of these cluster collisions \citep{Roettiger93,Loken95,Roettiger96}.  The X-ray surface brightness distribution is expected to become elongated along the merger axis and the central cD or D WAT host may become displaced from the large-scale center of the X-ray emission.  Shocks are expected to form initially along the merger axis and later perpendicular to this axis \citep{Roettiger96}.  During the merger the X-ray luminosity of the ICM is thought to increase \citep{SchMue93} due to an increase in ICM density.  Significant X-ray substructure should develop and persist within the ICM, and it is probable that cooling flows should be absent or disrupted \citep{McGlynn84, Gomez02}.  It is assumed that the ICM ram pressure during the cluster mergers is great enough to bend the WAT radio sources.  Bulk flow velocities are believed to remain above 1000 km s$^{-1}$ within $\sim$200 kpc of the cluster core for periods as long as 2 Gyr after the initial collision \citep{Loken95,Roettiger96}.  This time scale is much longer than the assumed lifetime of the radio sources (10$^7$-10$^8$ yrs). The bulk motion and increased density of the ICM may also result in ram-pressure stripping of the cooler interstellar medium of galaxies within the clusters \citep{Schindlerbook}.      

While Abell 1446 does not exhibit the tell-tale signs of a recent large-scale cluster-cluster merger, such as elongation of the X-ray emission or large pressure contrasts that would indicate a significant shock, there is possible evidence of a small-scale merger of a sub-cluster or group.  Along the line that bisects the opening angle of the WAT there appear to be enhancements in the two-dimensional maps of temperature and pressure as well as an excess of soft emission.  Such an excess, lying 20 kpc to the southeast of the central galaxy could possibly be a result of ram-pressure stripping of the central galaxy's interstellar medium by bulk flow of the ICM, the remnant cool core of the pre-merger cluster, or the core of an infalling group of galaxies ($\S 6.1.2$).     

Under the assumption that ICM ram pressure produced in cluster mergers shapes the WAT radio source, the line that bisects the WAT radio lobes should be aligned with the merger axis.  Simulations have shown that shocks propagate outward from the cluster center along the merger axis as the subcluster passes through the core \citep{Roettiger96}.  As seen in the two-dimensional maps of temperature and pressure there is possible evidence of a weak shock along the merger axis to the southeast of the WAT.  This region is roughly 50$\arcsec$(95 kpc) by 24$\arcsec$(46 kpc) with a temperature of $kT=5.93^{+2.06}_{-1.21}$ keV as compared to its adjacent cooler regions where $kT = 3.72^{+0.47}_{-0.33}$ keV (as discussed in $\S 4.3$).  In projection, the pressure map shows a region of higher pressure coincident with this region of increased temperature to the southeast of the WAT.  The region has a pressure roughly 1.6 times greater than the surrounding ICM pressure.  This region of enhanced pressure is coincident with the slight pressure irregularity in the radial profile discussed in $\S 5.1$.  We can determine an upper limit on the pressure ratio across the possible pressure discontinuity to be $\sim$1.9.  This corresponds to a Mach number of 1.3, or a merger shock velocity of $\sim$1460 km s$^{-1}$, determined from Rankine-Hugoniot jump conditions assuming a sound speed of 1120 km s$^{-1}$ for 4 keV gas.  In the classic model of WATs, such velocities are thought to be great enough to bend the radio sources into their swept back shape \citep{Burns86}. 

As the relative velocities of the ICM during the merger of a subcluster may shape the WAT radio source, it's possible the ICM conditions can also result in ram pressure stripping of the ISM contained within the cluster galaxies.  In X-ray observations and simulations it has been seen that elliptical galaxies can be stripped of their X-ray halos as they plow through the denser ICM near the cluster center \citep{Takeda84, Merrifield98, Stevens99, Drake00, TS01}.  It has been suggested that elliptical galaxies within merging clusters experience this ISM stripping as well \citep{Schindlerbook, Sakelliou05}.  The temperature of the X-ray halo is generally $\leq$ 1 keV, thus cooler than the surrounding gas of the ICM.  When the gas of the galactic ISM is stripped by the ram-pressure of the ICM, it is thought to leave a dense trail, or wake of cooler gas.  Such a wake was recently discussed in \citet{Sakelliou05} associated with the WAT 4C 34.16 and its host galaxy.  They argue that the wake, visible between the WAT lobes in a softness ratio image, was produced as the host galaxy fell into the cluster.  Given the ICM density of the cluster and the mass of the host galaxy, it was determined that a relative velocity of $\geq$ 1200 km s$^{-1}$ was necessary to produce the observed wake.  Through simulations, \citet{Acreman03} show that an infalling elliptical galaxy becomes mildly supersonic and loses its X-ray halo through ram pressure stripping during core crossing, reinforcing the wake claim of \citet{Sakelliou05}.  A soft X-ray image was created of A1446 to investigate the possible presence of a wake of stripped material behind the central galaxy, revealing substructure surrounding the AGN.  The energy range was restricted to 0.3-1.0 keV.  This along with the hard X-ray image (1.0-10.0 keV) is shown in Fig. 12.  A considerable excess in the soft image is present directly southeast of the AGN, between the lobes of the WAT.  It was determined that the excess is a significant enhancement over the average soft emission at the same radius from the AGN at the 2.85 $\sigma$ level.  A soft excess in this location is consistent with a wake of cooler galactic material having been stripped by the same ram-pressure that bent the WAT source.  In this case, the ram pressure is thought to have been supplied by a small-scale cluster merger.  We approximate the wake to be within a spherical volume of radius r$_{\rm wake}$ $\sim$ 20 kpc, with a density $\sim$ 1.2 n$_{\rm ICM}$ estimated from the excess soft emission. From this we calculate the mass of the wake to be M$_{wake}$ $\sim$ 10$^9$ M$_{\odot}$, which is comparable to the values for X-ray halos of elliptical galaxies (M = 10$^{6.5}$-10$^8$ M$_{\odot}$) determined in \citet{Sun07}.  While the temperature map (Fig. 8) shows that the wake region is not significantly cooler than the ambient medium, there is a substantial amount of projected cluster gas that may give this excess emission the appearance of having a higher temperature.  

Another explanation of this soft excess is that it may be the remnant cool core of the pre-merger cluster.  Within merging clusters it is believed that a pre-existing cooling core may be disrupted if the ram pressure of the infalling cluster exceeds the thermal pressure of the core \citep{Gomez02}.  The number of counts is not great enough to determine the density and temperature of the excess to significant accuracy.  As such, it is not possible to distinguish if it is the disrupted cool core of the pre-merger cluster, stripped interstellar material of the host galaxy, or the core of an infalling group of galaxies.

\subsubsection{Infalling Cool Clump}
It is possible that instead of a merger along the WAT axis, a group of galaxies, including gas cooler than the cluster's ICM, may be falling in from the north of the cluster.  This cooler gas would then be responsible for bending the radio source as it sinks in the ICM towards the cluster center.  The excess of emission directly to the north of the AGN (Fig. 5) may be such a clump of cool gas falling through the hotter ICM.  This clump can be seen in the adaptively smoothed image and residual image of the cluster as the bright filamentary structure extending northward from the cluster center.  The temperature map shows a region coincident with the northern filamentary clump that is cooler than the ambient cluster gas.  Following the method described in $\S 4.3$ the temperature of this region was determined to be $kT = 3.19^{+0.46}_{-0.43}$ keV while the surrounding regions have a temperature of $kT = 4.29^{+0.69}_{-0.56}$ keV.  The cool infalling material may be a primary factor in the bending of the northern lobe.  The presence of the cool clump in the path of the radio source may impede the normally straight large-scale flow of the energetic particles, forcing the structure to bend towards the east.  What appears to be a wake in the WAT axis merger scenario may either be the leading portion of this infalling cool clump that has already fallen through the cluster center or the bright edge of a bubble of compressed gas interacting with the northern lobe.  It has been argued that the symmetry of WAT sources rules out the possibility that the separate lobes are shaped by individual small-scale clumps in the ICM \citep{DeYoung91}.  While this scenario is able to explain the morphology of the northern radio lobe, it does not account for the similar bending direction of the southern lobe.           

\subsubsection{Line of Sight Merger}
We examined spectroscopically determined redshifts of galaxies within a 2 Mpc radius (at z=0.1035) of the center of Abell 1446 available in the SDSS database.  Within this radius 50 galaxies were found to lie $\pm$ 3000 km s$^{-1}$ from the mean at $\sim$ 31000 km s$^{-1}$.  Figure 13 (left panel) shows the velocity distribution of these 50 galaxies presumed to be members of A1446.  While the majority of the galaxies are within 1000 km s$^{-1}$ of the central peak, consistent with the velocity distribution of a moderately relaxed cluster, a secondary peak of galaxies is offset from the central distribution by $\sim$ 1500 km s$^{-1}$.  Such a velocity offset is consistent with a merger of a subcluster occurring, at least partially, along the line-of-sight.  The right panel of Figure 13  shows the positions on the sky of the members of A1446 with SDSS redshifts.  The galaxies are divided into two subsamples: lower velocity galaxies are shown as triangles while higher velocity galaxies are shown as circles.  The majority of the lower velocity galaxies (9/10) roughly occupy the nothwest quadrant of the plot.  These galaxies are also generally distributed along the line that bisects the WAT, and possibly associated with the northern filamentary excess shown in Fig. 5.  The similarly low velocities of these galaxies, the coincidence of sky positions, and the bending direction of the WAT are all consistent with a merger of a subcluster occurring from the northwest to the southeast with a significant component ($\sim$ 1500 km s$^{-1}$) along the line-of-sight.  Previously published velocity measurements of the galaxies within Abell 1446 \citep{Pinkney00} do not show this bi-modal velocity distribution.  From a region of the sky 3 x 3 Mpc$^2$, they obtained spectra for 34 cluster members yet only 3 were within the lower velocity range depicted by the triangles in Fig. 13.

While there is evidence for a bi-modal velocity distribution (SDSS) in Abell 1446, with the galaxies forming the secondary peak lying along the line that bisects the WAT, the sample is relatively small.  Abell 1446 is of moderate richness (Abell richness class 2, 80-129 galaxies between m$_{\rm 3}$ and m$_{\rm 3}$+2 \citep{Abell58}) thus 50 galaxies is not a complete sampling of the cluster velocity structure.  Additional spectroscopic observations are required to increase the number of spectra to N $\sim$ 100-200 to further investigate this possible line-of-sight merger.    

\subsection{Radio Morphology and Galaxy Velocity Estimates}
When they were first observed, it was suggested that WATs could be created by the swift orbital motion of the host galaxy through a relaxed ICM \citep{OR76} as is often seen with Narrow Angle Tail (head-tail) galaxies.  NATs are usually smaller cluster members that are orbiting the cluster center at great velocities farther from the core than WAT sources.  These greater orbital velocities produce significant ram pressures allowing the lobes to be bent into a narrow angle.  As mentioned in \citet{Burns81}, the high velocity NAT explanation fails for WATs as D and cD type galaxies are expected to lie at the center of their clusters with low peculiar velocities ($v_{g}< 300$ km s$^{-1}$).  

It must be noted that while the direction of NAT tails has been found to be consistent with random orbital motions of galaxies within their host clusters \citep{Odea87},
some galaxy clusters host multiple NATs with similarly oriented tails [e.g. Abell 119, \citet{Feretti99}].  This along with the finding of \citet{Bliton98} that NAT hosting clusters exhibit a significant degree of X-ray substructure suggests that cluster mergers contribute in part to the formation of these head-tail sources.

Using an \emph{Einstein} image of A1446 along with 4.8 GHz VLA observations, \citet{BB82} examined 1159+583.  They invoked a nonrelativistic hydrodynamic flow model to explore the radio plasma properties required to reproduce the observed bending.  They assume the host galaxy to be moving through the ICM with a velocity $\sim$ 200 km s$^{-1}$ and argue that the radio lobes are rising away from the dense cluster center.  Within their model, buoyant and dynamic forces contribute equally to the shaping of the WAT.  Euler's equation can then be used in the form:   

\begin{equation}       
\frac{\rho_{r}v^{2}_{r}}{r_{c}}=\frac{\rho_{\rm ICM}v^{2}_{g}}{r_{r}} + \nabla P      
\end{equation}

\noindent where $r_{r}$ is the radius of the lobe, $\rho_{r}$ is the mass density of the lobe, $v_{r}$ is the velocity of the plasma within the lobe, $r_{c}$ is the radius of curvature of the source, $v_{g}$ is the velocity of the host galaxy relative to the ICM, $\rho_{\rm ICM}$ is the density of the ICM, and $P$ is the ICM gas pressure.  The last term is due to buoyancy.  They assume the relativistic electrons must travel at $>$ 10$^4$ km s$^{-1}$ in order to travel between the galaxy core and the end of the lobe within their synchrotron lifetime [as computed in \citet{BOR79}].  Given this velocity and the model above they determined the density of the plasma within the lobe to be $<$ 2 $\times$ 10$^{-5}$ cm$^{-3}$.  

In a survey of 11 WAT radio sources (including 1159+583), \citet{ODon93} also tested the possibility of source bending through slow galactic motion.  Their first model, the adiabatic model, assumed no acceleration of energetic particles outside the radio core.  Such a model is similar to that used in \citet{BB82}.  With no \emph{in situ} particle acceleration in the lobes, \citet{ODon93} found that the sources were easily bent by the galaxy's slow motion through the ICM.  Although the lobes were found to bend easily within this model, they note that there were serious inconsistencies between the travel time of the particles down the lobe and their spectral age.  Their second model, the kinetic model, is similar to that in \citet{Eilek84} and assumes that particles are accelerated outside the core by fluid processes.  This requires that the radio luminosity is supplied by a fraction of the kinetic energy of the particles flowing down the lobe.  The condition used to determine the internal density required to produce the observed luminosity of the lobes, given the plasma flow velocity is:  

\begin{equation} 
L_{\rm rad}=\frac{\epsilon\pi r^{2}_{r} \rho_r v^{3}_{r}}{2}
\end{equation}

\noindent A form of Euler's equation (eq. 1 without the buoyancy term) is then used to determine the velocity of the host galaxy relative to the ICM necessary to produce the bent WAT morphology (Eq. 3) as:

\begin{equation} 
v_{g}=\left(\frac{2 L_{\rm rad}}{\epsilon \pi \rho_{\rm ICM} v_r r_c r_r}\right)^{1/2}       
\end{equation}

\noindent where $L_{rad}$ is the total radio luminosity of the lobe and $\epsilon$ is the radiative efficiency (0.001-0.1) which measures the conversion from kinetic to relativistic particle energy \citep{Eilek79}.  Given the more recent 1.4 GHz and 4.8 GHz VLA data, \citet{ODon93} determined a significantly different radio source geometry than that used in \citet{BB82}.  While \cite{BB82} used the radius of curvature and radio lobe radius values of 200 kpc and 12 kpc respectively, \citet{ODon93} used 16 kpc and 1 kpc.  Such different input parameters within this kinetic model significantly alter the result. Assuming that the initial particle velocity is $v_r = 0.2c$, decelerating to 0.11$c$ at the bend, and that $\epsilon < 0.01$, \citet{ODon93} found that a galaxy velocity of $v_g > 3400$ km s$^{-1}$ was necessary to reproduce the observed bending of 1159+583.  This value is calculated assuming n$_{\rm ICM}$ = 10$^{-3}$ cm$^{-3}$.  With the new Chandra data it was determined that n$_{\rm ICM}$ $\sim$ 0.0025 cm$^{-3}$, so that a galactic velocity of 1800 km s$^{-1}$ is required to bend the radio source.

Using VLA archival data of 1159+583, we re-examined the properties of the bent source.  The two different geometries chosen by \citet{BB82} and \citet{ODon93} appear to be dependent on the scale over which the ICM/radio interaction is believed to be occurring.  \citet{ODon93} assume that the bending occurs predominantly at a position just downstream of the radio hotspots over a relatively short distance.  \citet{BB82} instead assume the bending occurs across the entire source and as such measure a larger lobe radius and radius of curvature.  Within the model that attributes the bending of the radio source to relative bulk flow of the ICM, the ram pressure of the gas is believed to act on the entire source.  As such, we will take the radius of curvature of the large-scale structure of 1159+583 as opposed to localized bends near the hotspots. A 1.4 GHz map displays a radius of curvature $\sim$ 100 kpc with a lobe radius of $\sim$ 7 kpc, which is of the order of the \citet{BB82} values determined from lower resolution radio data.   

A spectral index map of 1159+583 was obtained between frequencies near 1.4 GHz and 4.8 GHz by convolving each data set to a common circular beam size of 1.61$\farcs$  The resulting spectral index map was blanked at the 3$\sigma$ level of each of the input frequencies and balanced between natural and uniform weighting for better sensitivity to diffuse emission.  This can be seen in Figure 13.  The spectral index, $\alpha$, of the northern jet is 0.6, where $\rm S_\nu \propto \nu^{-\alpha}$.  The southern jet was not detected.  From the spectral index map it can be seen that $\alpha$ steepens with increasing distance from the core with occasional regions of spectral flattening.  At these locations of spectral flattening, it is likely that a fraction of the kinetic energy is transferred to the relativistic particles \citep{Muxlow91}.  This re-energizes the particles and prevents the spectrum from showing significant signs of aging as the distance from the central engine increases.  To examine this more closely  a radial profile of the spectral index of the southern lobe was calculated from the 1.4 GHz and 4.8 GHz maps (Fig. 14).  While there is an overall steepening of the spectral index along the length of the southern lobe, the slope remains relatively shallow.  This region of nearly constant spectral index between 15$\arcsec$ and 40$\arcsec$ is consistent with \emph{in situ} particle acceleration within the radio plasma \citep{Parma99}. Assuming this \emph{in situ} particle acceleration, the global luminosity condition along with the form of Euler's equation presented above (Eq. 3) can be used to estimate the galaxy velocity necessary to produce the bent tail morphology.  A significant buoyant force acting on the radio lobes, included in the calculations of \citet{BB82}, would result in the radio plasma bending radially away from the cluster center as it rises in the ICM.  Given the orientation of the radio source relative to the large-scale X-ray emission it is difficult to determine the effect of buoyancy on the lobes (acting with or against ram-pressure). Thus the $\nabla P$ term in Eq. 1 is not included.

The luminosity condition (Eq. 2) is based upon the argument that some fraction ($\epsilon$) of the kinetic energy of the bulk flow is converted to relativistic particle energy resulting in the observed synchrotron emission.  This value is not well known and has generally been assumed to be between 0.001-0.1.  \citet{Birzan04} attempted to constrain this value by examining a sample of 16 galaxy clusters that have clear decrements in the X-ray emission coincident with the lobes of cluster center radio sources.  It is assumed that this decrement in X-ray emission is a result of the relativistic plasma of the radio sources carving out cavities in the ICM.  The energy necessary to evacuate the cavities can be estimated from the product of the cavity volume and overlying ICM pressure.  Assuming a standard time scale over which the cavity is created ($5 \times 10^7 - 10^8$ yr), a kinetic luminosity can then be estimated.  The efficiency ($\epsilon$) is then the ratio of total radio luminosity to kinetic luminosity.  \citet{Birzan04} still found that $\epsilon$ ranged in values between 0.001-0.1 for their sixteen sources.  This method results in a rough estimate of a radio source's kinetic efficiency. As such we carried out a similar analysis on A1446 and 1159+583 to compute $\epsilon$ rather than simply choosing an arbitrary value.  There is a clear deficit of emission coincident with the northern radio lobe.  A time scale of $5 \times 10^7$ yr was assumed which is consistent with the time required for the cavity to rise the projected distance from the radio core to its present location at the speed of sound.  This results in the kinetic efficiency estimated as $\epsilon \sim 0.04$.       
  
The total luminosity of the source, which is dominated by the lobes, was found to be 3.4 $\times$ 10$^{42}$ ergs s$^{-1}$.  This was determined using the 1.4 GHz flux as a reference and assuming a power law of spectral index $\alpha$ = 0.85, taken as the average of the spectral index map. The total flux was calculated by integrating between $10^7$ and $10^{11}$ Hz.  From the total flux the total radio luminosity was determined accounting for its dependence on spectral index and redshift.  With values for $\rho_{\rm ICM}$, $r_c$, $r_r$, $\epsilon$ determined as $4.2 \times 10^{-27}$ g cm$^{-3}$, 100 kpc, 7 kpc, and $0.04$, respectively, the only unknown is $v_r$.  Previous estimates for the conditions within the lobes of WATs \citep{Sakelliou96,Smolcic06} have used observed effects of buoyancy to constrain internal parameters.   Their sources show apparent bends due to ram pressure close to the host galaxy, but as the distance from the center increases the lobes bend towards the opposite direction.  This is the position where buoyancy balances ram pressure and is then used to constrain the internal density or flow velocity within the source.  As discussed above the impact of buoyancy on the morphology of 1159+583 is unclear.  There is no evident transition point where buoyancy balances ram pressure.  While \citet{Sakelliou96} and \citet{Smolcic06} determined internal velocities of their WAT sources to be $v_{lobe} <$ 0.01c and $v_{lobe} \sim$ 0.045c respectively, the flow velocity within the lobes of 1159+583 remains poorly constrained.  

Using a beaming analysis, the plasma flow velocities down the well collimated jets of WAT sources have been estimated to be between (0.3-0.7)$c$ \citep{Jetha06}.  Flow velocities within the lobes are less restricted but may be within the range of (0.005 - 0.2)$c$.  From Eq. 3, this range of flow velocities results in a broad range of galaxy velocities ($v_g$ $\sim$ 200-1000 km s$^{-1}$) capable of bending 1159+583.  It must be kept in mind that the value for kinetic efficiency ($\epsilon = 0.04$) is a rough estimate as well which contributes to the uncertainty in the galaxy velocity.  Figure 15 shows a plot of plasma flow speed down the lobes versus kinetic efficiency.  The diagonal line represents the combinations of plasma flow speed and kinetic efficiency required to bend the lobes into their observed shape given a galaxy velocity of 300 km s$^{-1}$ (the upper limit for systematic galaxy motion about the cluster center \citep{Burns81, QL82,Malumuth92}).  Any combination of efficiency and flow speed that lies above this line is consistent with slow galaxy motion sufficiently bending the lobes.  Any combination that falls below the line requires galaxy velocities greater than 300 km s$^{-1}$ to bend the lobes, consistent with merger induced bulk flow of the ICM.  Assuming our estimated kinetic efficiency of $\epsilon \sim 0.04$, a plasma flow speed $<$ 0.06$c$ would require galaxy velocities relative to the ICM consistent with conditions expected at the site of a cluster merger.  Recently, \citet{Hardcastle05} argued that the slow galactic motion model can explain WAT bending as long as the plasma flow in the lobes is faster and significantly less dense than the typical values used in previous models. From the global luminosity condition (taking $\epsilon \sim 0.1$), and assuming flow speeds down the lobe $>$ 0.08$c$ they calculate that galactic velocities as slow as 100-300 km s$^{-1}$ are sufficient to bend the lobes of the WAT radio source 3C 465.  With the kinetic efficiency in 1159+583 of roughly $\epsilon \sim 0.04$, we also find that flow speeds $> 0.06c$ result in lobe bending by relative velocities consistent with systematic galaxy motion about the cluster center.  While the kinetic efficiency is moderately constrained the plasma velocity is not. As such, within reasonable limits on the two values, both systematic galaxy motion and merger induced bulk flow of the ICM are possible explanations for the shaping of 1159+583.

\subsection{Pressure Balance between Radio Lobes and ICM}
While it is commonly assumed that there should be pressure balance between the ICM and radio lobes, \citet{Gomez97}, using \emph{ROSAT} and VLA data \citep{ODon90}, state the radio source 1159+583 is a dramatic counterexample for the equilibrium between thermal ICM pressure and the radio pressure.  They find an ICM pressure from the \emph{ROSAT} data that is significantly lower than the radio equipartition pressure.  This striking imbalance between the ICM pressure and radio pressure is not confirmed by our higher spatial resolution \emph{Chandra} observation.  This discrepancy is a result of at least two factors within \citet{Gomez97}.  These include their lower temperature of $2.6^{+1.3}_{-0.70}$ keV as well as their lower density, determined from the beta model fit to the surface brightness profile. The average minimum pressure of the radio lobes assuming equipartition is $\sim$ 3 x 10$^{-12}$ dyn/cm$^{2}$ \citep{ODon93}, while the ICM pressure ranges from 5.3 x 10$^{-11}$ dyn cm$^{-2}$ at the cluster center to 3.5 x 10$^{-11}$ dyn cm$^{-2}$ at the radius of lobe termination.  Thus, the lobes depart from the minimum energy condition or there is an additional component within the lobe regions supplying pressure, such as very hot, diffuse, thermal gas.  Similar results have been found previously for radio lobes of FR I sources within clusters [e.g. \citet{Morganti88,Killeen88,Feretti90,Feretti92}].

\section{CONCLUSION}
\emph{Chandra} spectra of Abell 1446 reveal that the intracluster gas is isothermal when averaged over circular annuli to a radius of 400 kpc with relatively smooth profiles of density and pressure.  The cooling time of the gas is greater than a Hubble time at all locations explaining the absence of an ongoing cooling flow.  The cluster exhibits no radial abundance variation and its $L_{X}-T$ relation appears to fall along the trend derived in \citet{Wu99}.  The gas mass to gravitational mass fraction converges to $\sim$0.1 at large radii as is seen in the majority of clusters \citep{Allen04}.

With no significant elliptical elongation of the cluster's large-scale ICM distribution and no large discontinuities in the radial profiles of temperature or pressure, it is likely that Abell 1446 has not experienced a large cluster-cluster merger in its recent history.  Abell 1446 may be experiencing small-scale mergers with subclumps or galaxy groups resulting in the presence of temperature and pressure substructure along the line that bisects the WAT.  If the excess directly to the southeast of the AGN is a wake, it would be expected that this material was stripped from its host galaxy by the infalling gas.  It is also possible that this soft excess is instead the remnant cool core of the pre-merger cluster or the remnant core of a group infalling from the north.  Additionally, spectroscopically determined redshifts of the member galaxies suggest that some portion of a merger ($\sim$ 1500 km s$^{-1}$) may be occurring along the line-of-sight.  

The plasma flow speed down the radio lobes of 1159+583 is poorly constrained.  As the two leading explanations for WAT bending are merger induced bulk flow of the ICM and systematic galaxy motion relative to the cluster mean, neither can be ruled out given the uncertainties described above.  Using our value of $\epsilon \sim 0.04$, velocities in the lobes $>$ 0.06c are required to bend the radio source without invoking a merger.  These flow velocities of particles down the radio lobes are generally higher than the values traditionally assumed but consistent with those used in \citet{Hardcastle05}.  

Support for this work was provided by the National Aeronautics and Space Administration through Chandra Award numbers G04-5148X,  G05-6126X, and GO5-6137X issued by the Chandra X-Ray Center, which is operated by the Smithsonian Astrophysical Observatory for and on behalf of NASA under contract NAS 8-39073.  Basic research in radio astronomy at the Naval Research Laboratory is supported by 6.1 base funding.  Elizabeth L. Blanton was supported by a Clare Boothe Luce Professorship

\bibliography{apjmnemonic,references}
\bibliographystyle{apj}

    \begin{figure}
\epsscale{1.1}
      \plotone{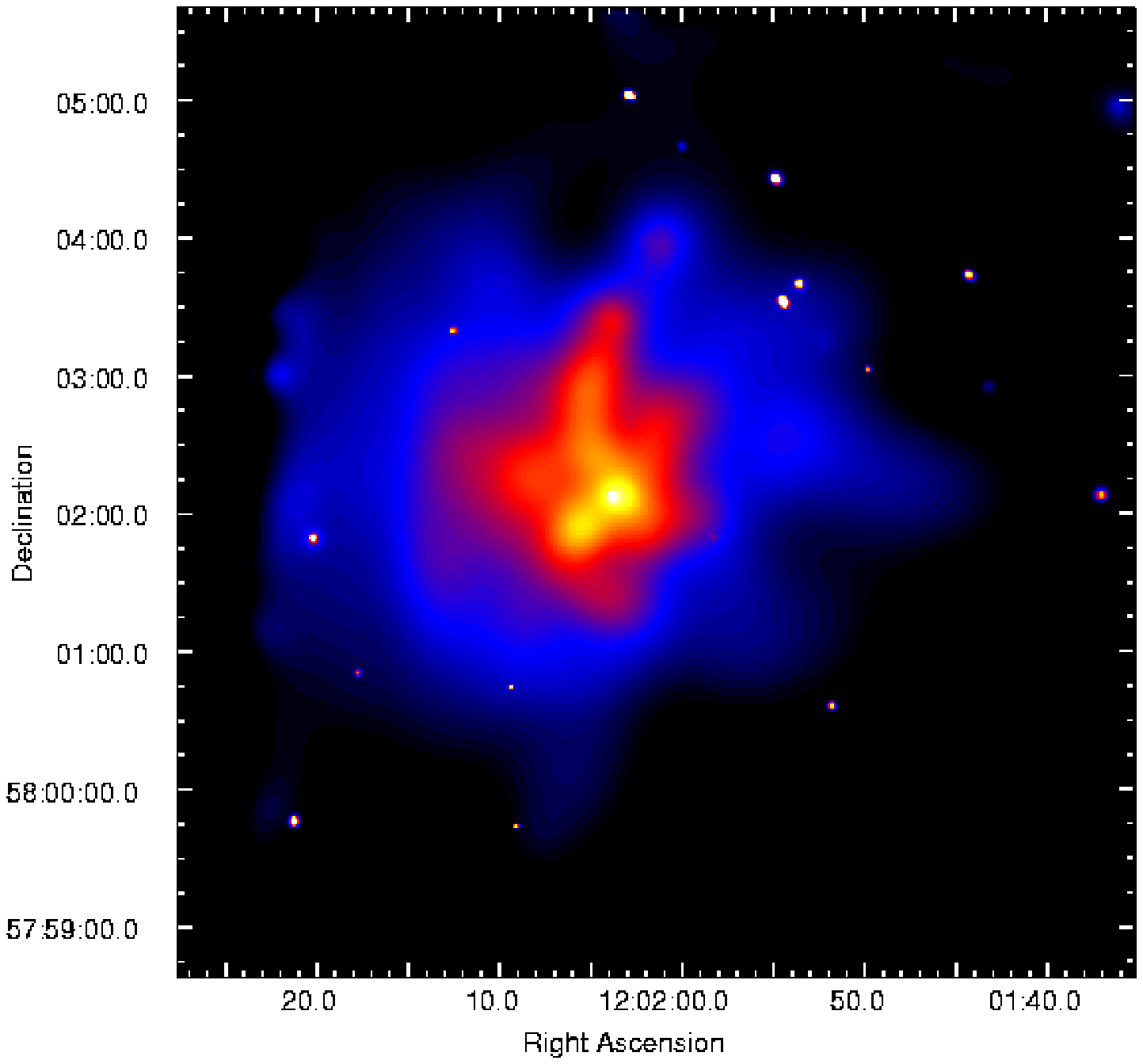}
      \caption{\label{fig:sed} An adaptively smoothed Chandra ACIS-S3 image of the 800 $\times$ 800 kpc$^2$ region of the galaxy cluster Abell 1446.  The image has been corrected for background and exposure.  On larger scales the cluster is approximately circularly symmetric but the inner 100$\arcsec$ display anisotropies in the distribution of the gas.}

 \end{figure}

    \begin{figure}
\epsscale{0.9}
      \plotone{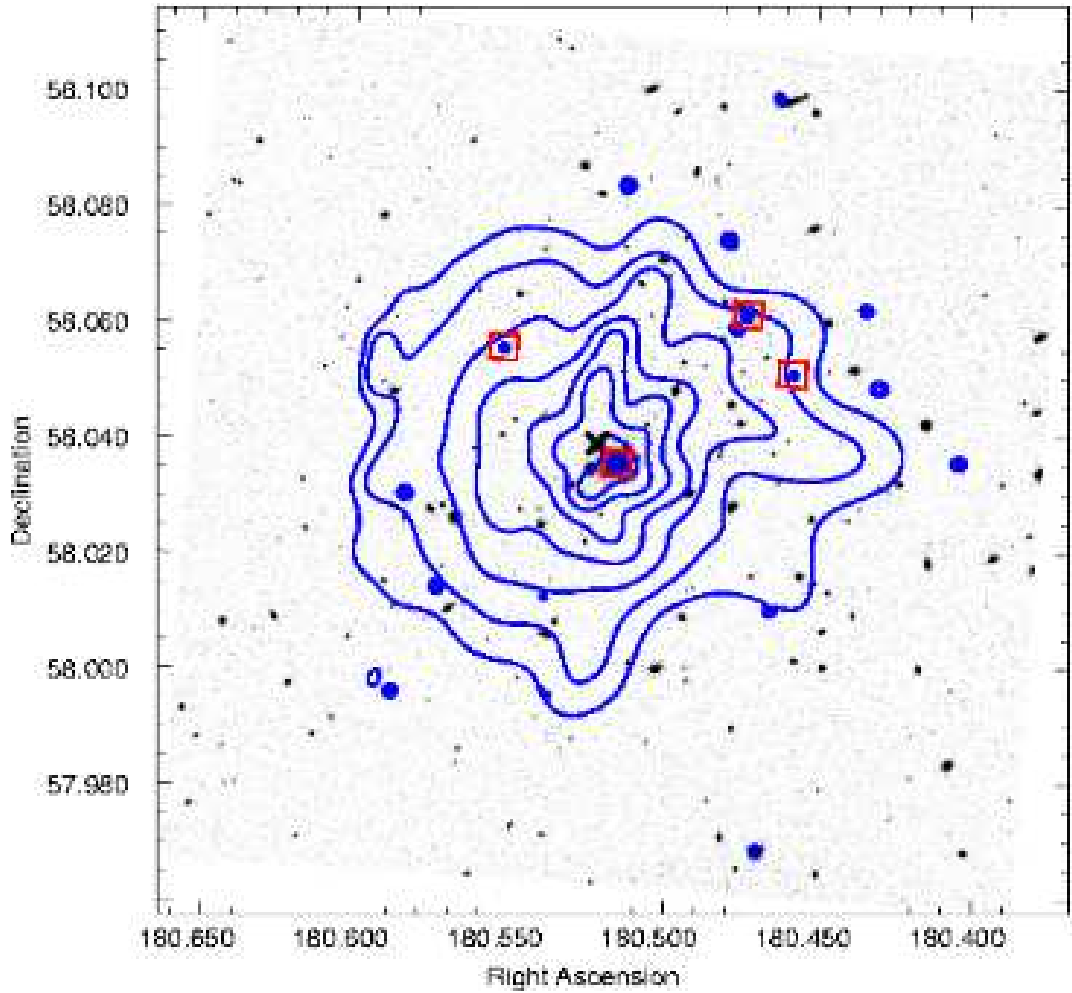}
      \caption{\label{fig:sed}SDSS R-band image of Abell 1446 with overlays of X-ray contours of adaptively smoothed emission obtained with Chandra (Figure 1).  The contours are logarithmically spaced between 0.17 cts arcsec$^{-2}$ and 2.6 cts arcsec$^{-2}$. The lowest contour is at background(blank sky) + 2$\sigma$.  The peak in the X-ray emission corresponds with the large elliptical galaxy which is the host of the WAT approximately at the center of the cluster.  The `X' marks the center of the large-scale diffuse emission.  The red squares indicate X-ray point sources identified using the \emph{wavdetect} algorithm which are coincident with optical sources in the USNO A2.0 catalog.}

 \end{figure}

    \begin{figure}
	\epsscale{0.85}
      \plotone{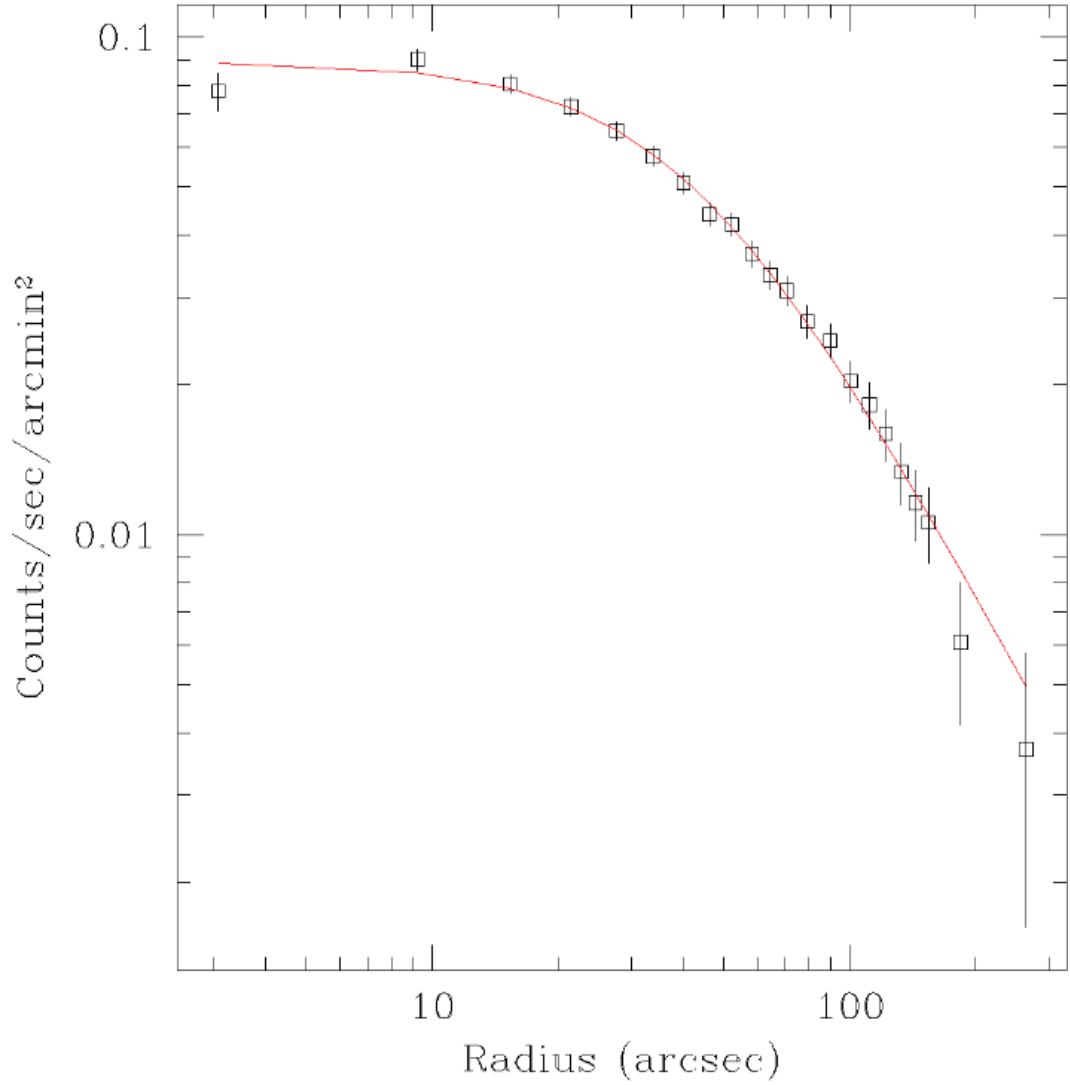}
      \caption{\label{fig:sed} Surface brightness profile (0.3-10.0 keV) and single $\beta$-model fit of the ICM.}

    \end{figure}

    \begin{figure}
	\epsscale{1.0}
      \plotone{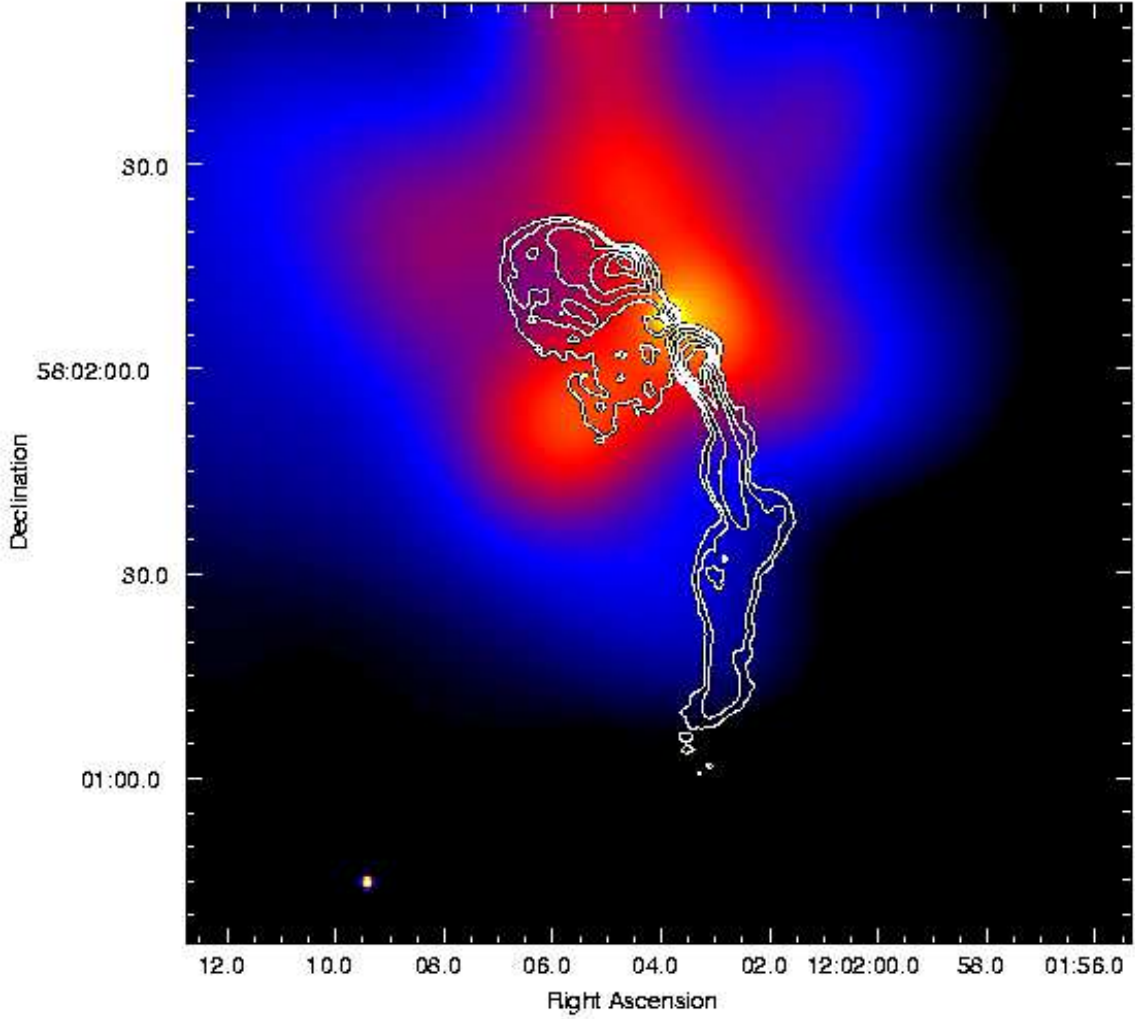}
      \caption{\label{fig:sed} A 200 $\times$ 200 kpc region of the cluster center overlaid with 1.4 GHz radio contours (A+C configuration VLA archival data) of the wide angle tail radio source 1159+583.  The contours shown are 5,10,30,60,100,156, and 223 times the local rms.  On close visual inspection it appears that the northern lobe has carved out a cavity within the cluster gas.  An excess of emission is seen between the lobes.}

    \end{figure}

    \begin{figure}
      \plotone{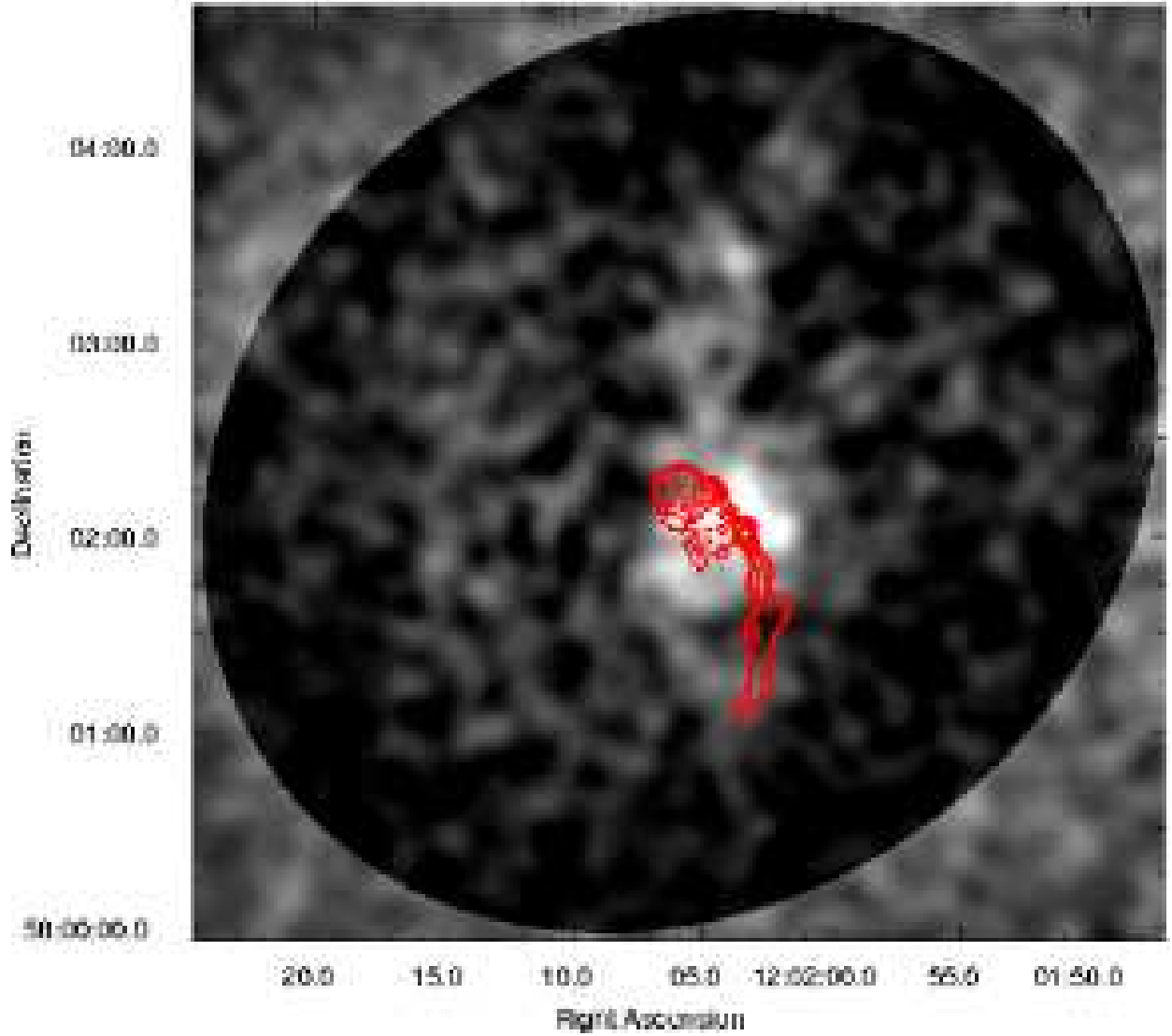}
      \caption{\label{fig:sed} An image of excess X-ray emission created by subtracting an isophotal model from a 4$\arcsec$ Gaussian smoothed image of the cluster.  The model was weighted by 0.7 to avoid over subtraction.  Radio contours are the same as in Fig. 4.}

    \end{figure}

   \begin{figure}
      \plotone{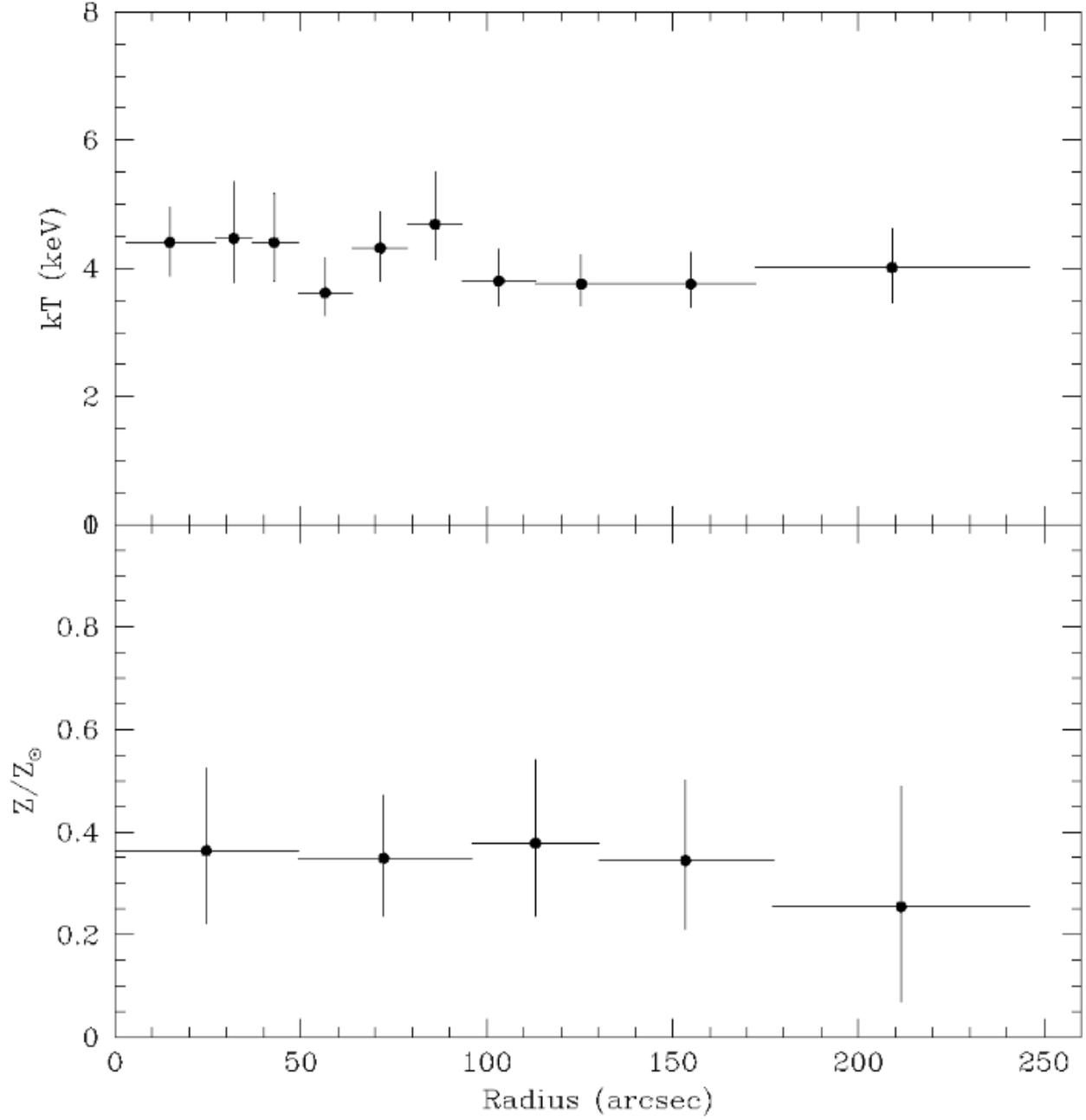}
      \caption{\label{fig:sed} Profiles of temperature and chemical abundance.  Both profiles show no strong dependence on radius.}
    \end{figure}

    \begin{figure}
	\epsscale{0.85}
      \plotone{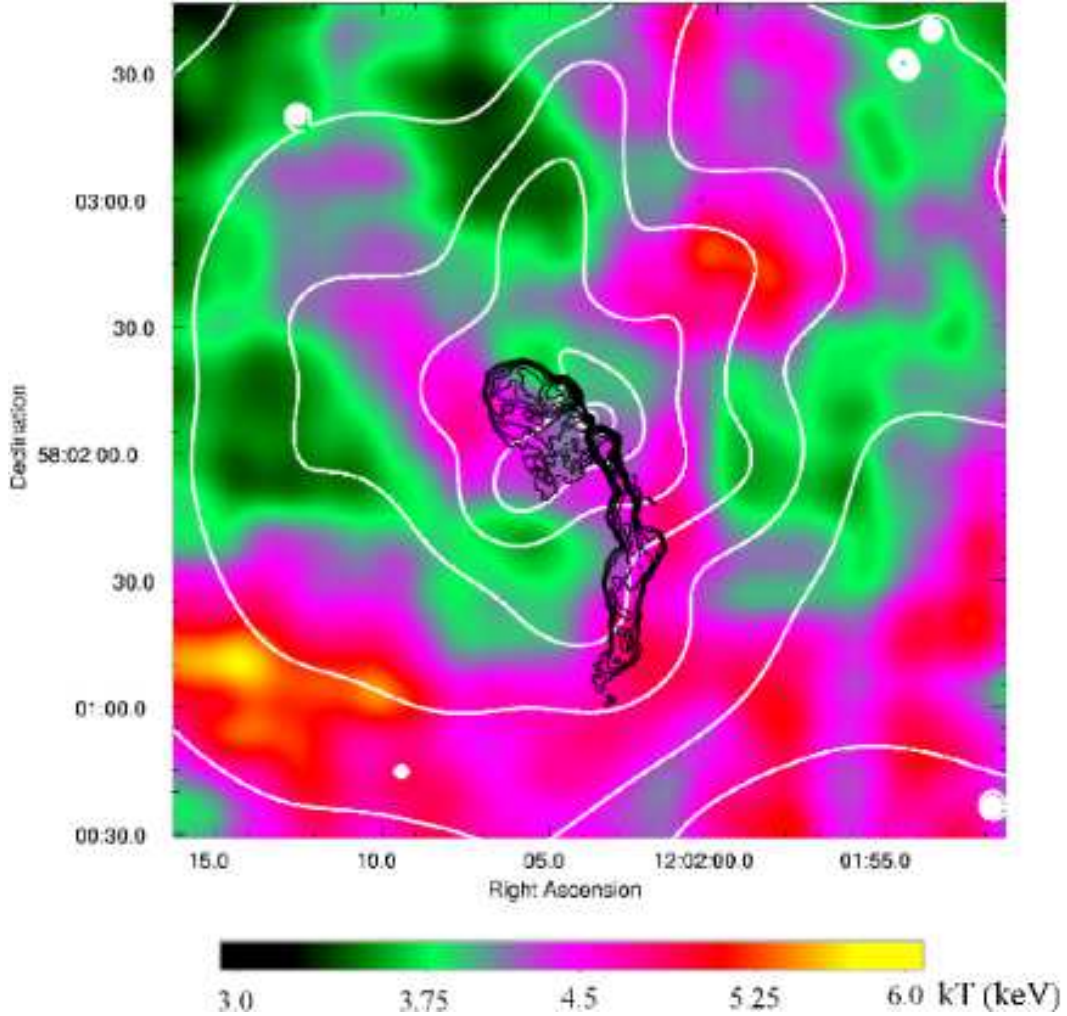}
      \caption{\label{fig:sed} Temperature map of the cluster center with contours of the smoothed \emph{Chandra} emission (white) and 1.4 GHz radio emission (black) superposed.  A region of significantly hotter gas can be seen to the southeast of the AGN along the line that bisects the WAT.  In addition, the excess of emission to the north seen in Fig. 5 may be identified as the cooler region directly north of the cluster center. }

    \end{figure}

    \begin{figure}
      \plotone{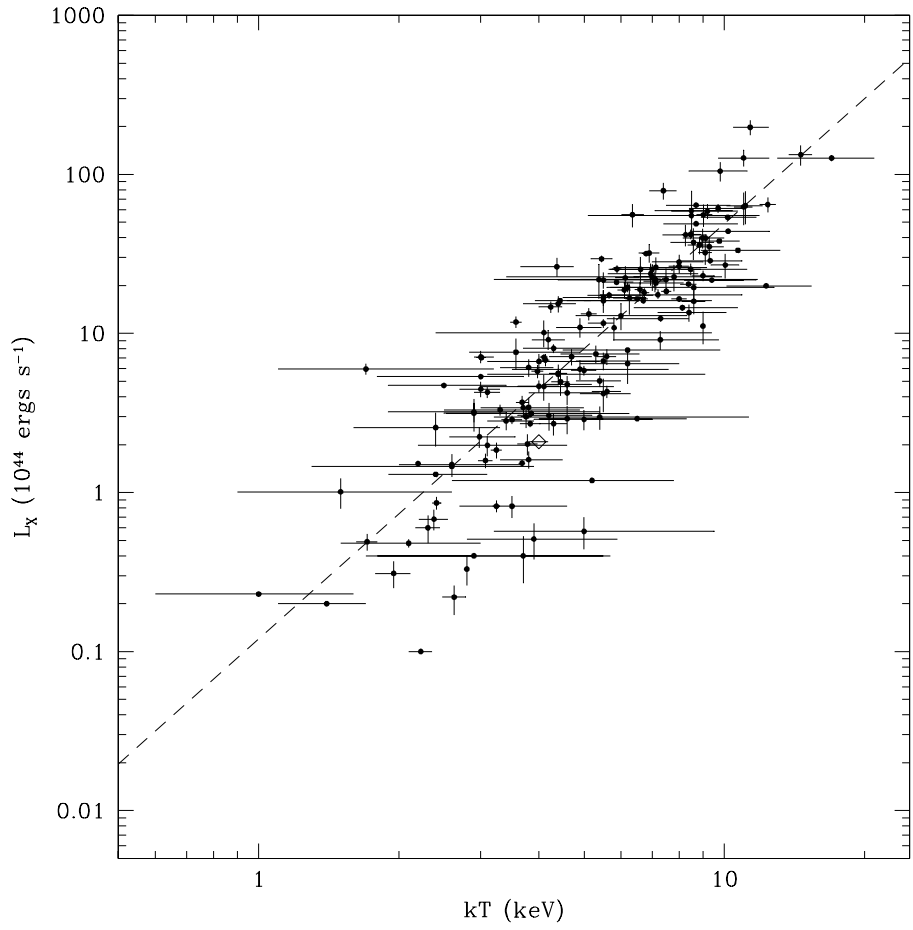}
      \caption{\label{fig:sed} $L_X-T$ relation plot from \citet{Wu99}.  Abell 1446 is denoted by the diamond.  The relation appears to fall roughly along the trend.}

    \end{figure}

   \begin{figure}
      \plotone{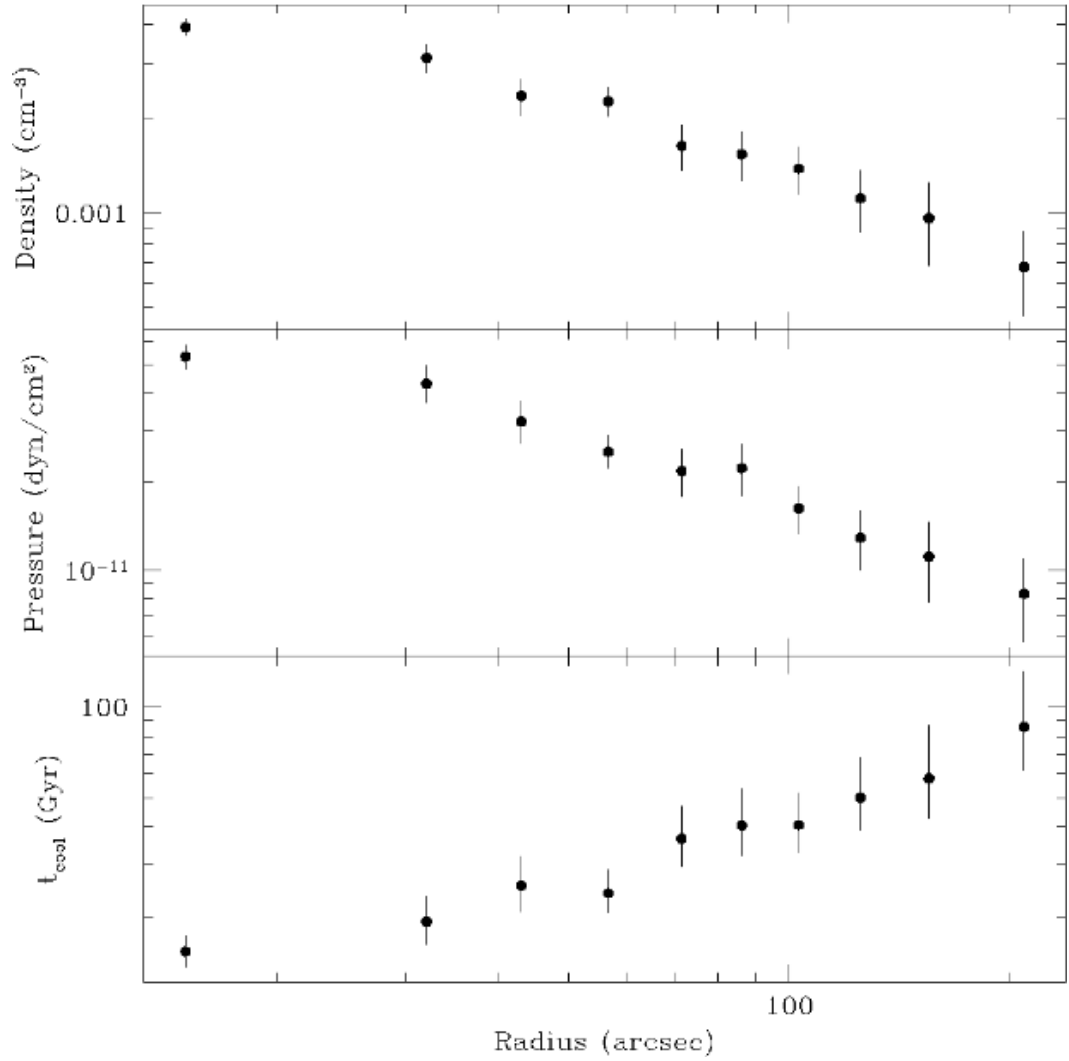}
      \caption{\label{fig:sed} Profiles of electron number density, pressure, and cooling time.  There is no evidence for a sharp change in pressure from a significant shock as would be associated with a large-scale cluster merger.}
    \end{figure}

   \begin{figure}
   \epsscale{0.75}
      \plotone{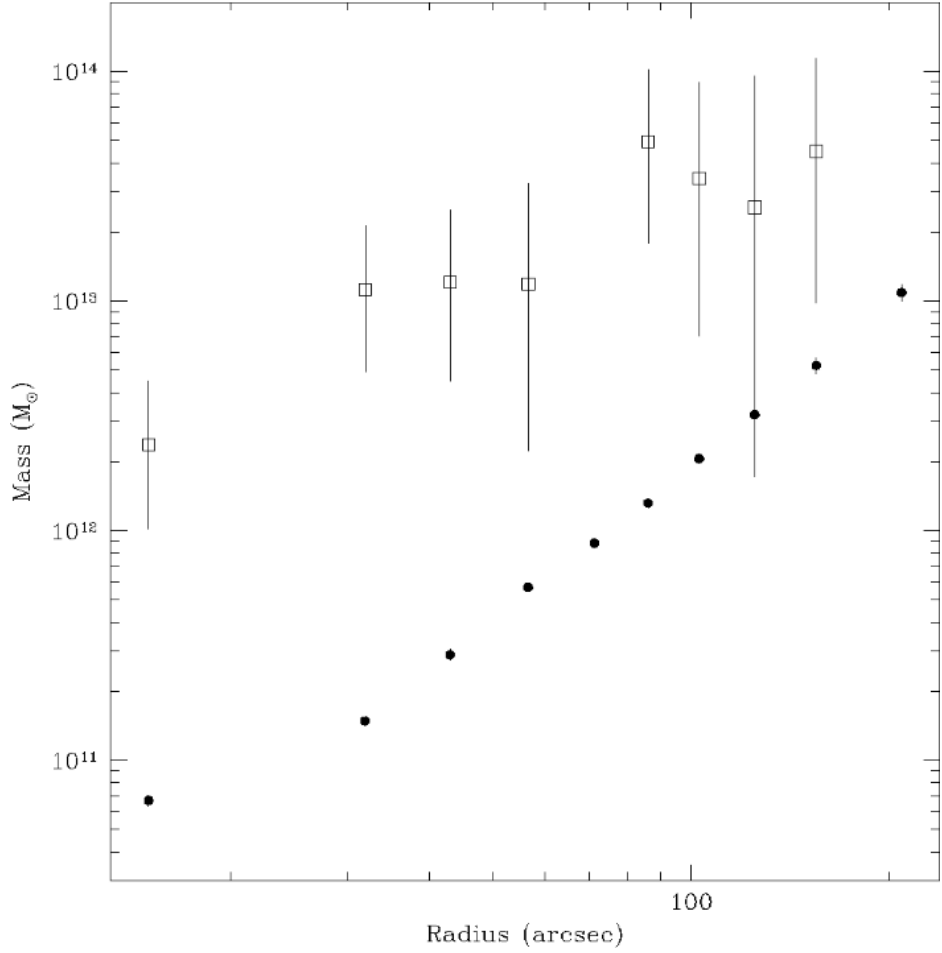}
      \caption{\label{fig:sed} Profiles of gas (circles) and gravitational (squares) mass. As the radius increases, the gas mass fraction converges towards 0.1 as is seen in other galaxy clusters.}
    \end{figure}

    \begin{figure}
	\epsscale{0.85}
      \plotone{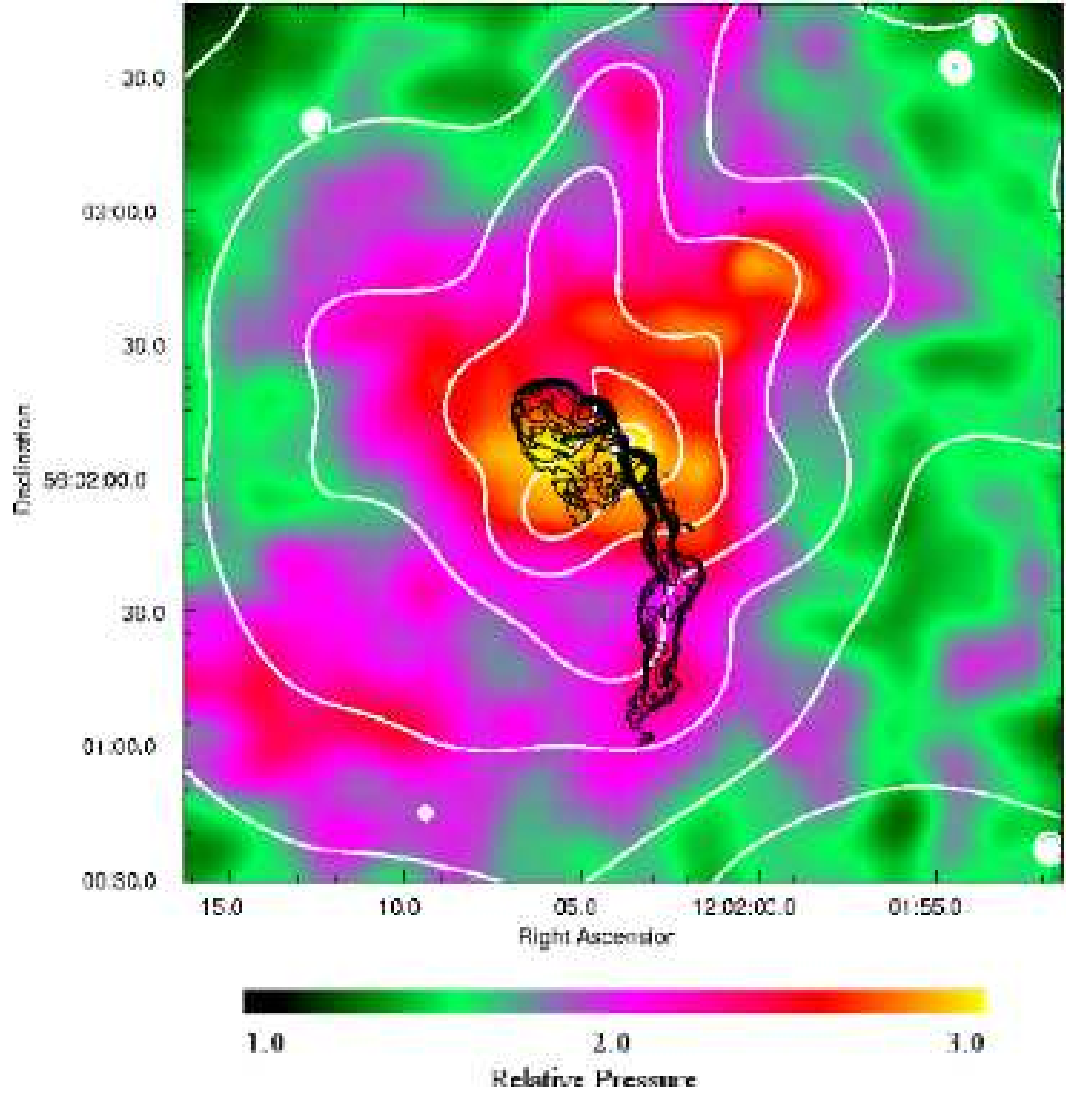}
      \caption{\label{fig:sed} Pressure map of the cluster center with contours of the smoothed \emph{Chandra} emission (white) and 1.4 GHz radio emission (black) superposed.  Pressure substructure can be seen along the line that bisects the WAT.  A region of high pressure is coincident with the southern radio lobe.}

    \end{figure}

    \begin{figure}
      \plotone{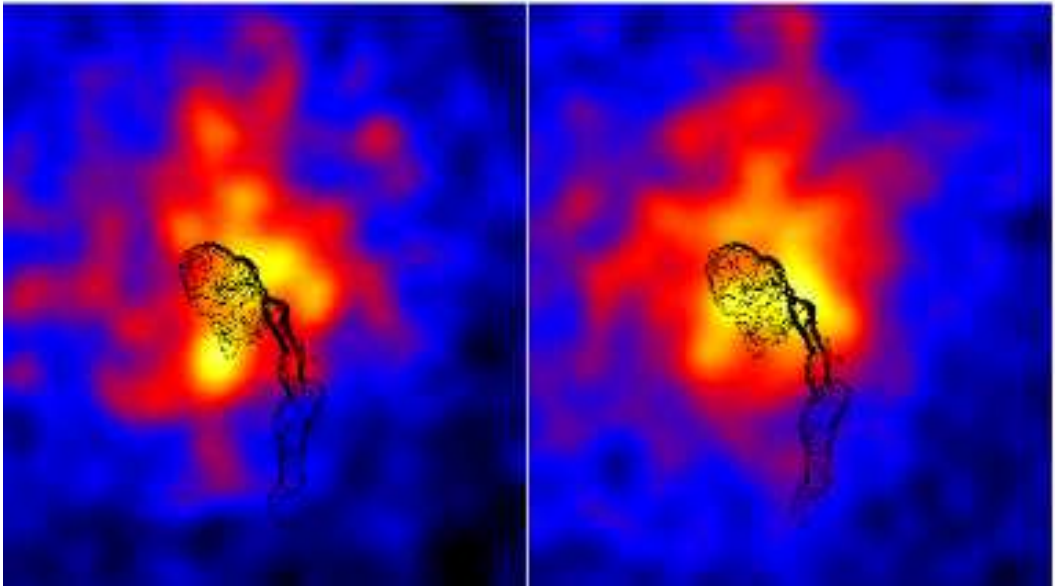}
      \caption{\label{fig:sed} Soft [E = 0.3-1.0 keV](left) and hard [E = 1.0-10.0 keV](right) images of the central region of Abell 1446.  The wake candidate can be identified in the soft image as the excess emission to the southeast of the AGN.}
	\end{figure}

    \begin{figure}
\epsscale{1}
      \plotone{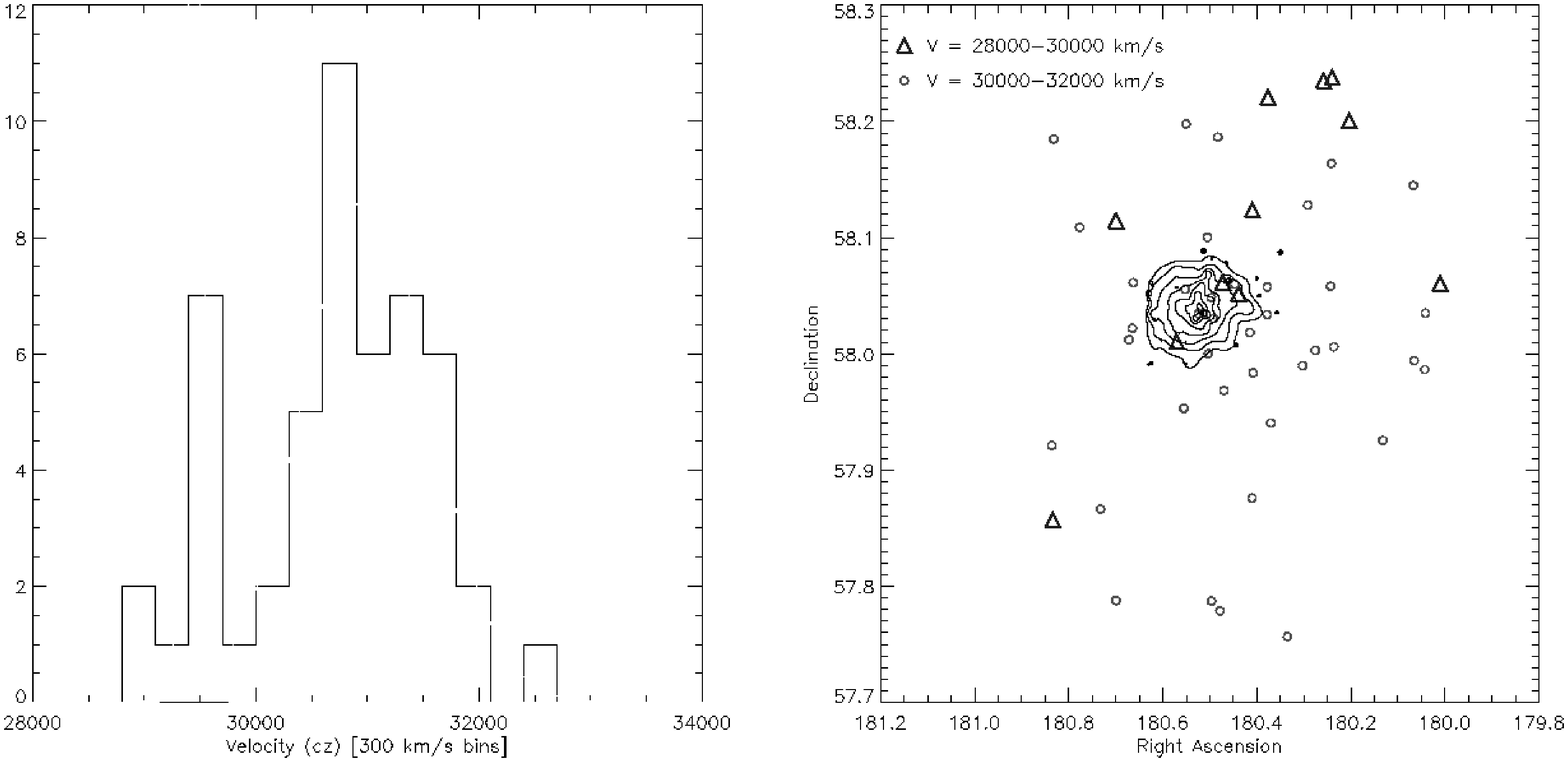}
      \caption{\label{fig:sed} \emph{left:} Velocity distribution of 50 cluster members of A1446 with spectroscopically determined redshifts from SDSS.  A secondary velocity peak is offset from the central distribution by $\sim$ 1500 km s$^{-1}$. \emph{right:} Distribution on the sky of the cluster members with SDSS redshifts.  The data is divided into two samples defined by the velocity separation of the primary and secondary populations.  Lower velocity galaxies with v = 28000 - 30000 km s$^{-1}$ are shown as triangles while those with higher velocites v = 30000 - 32000 km s$^{-2}$ are shown as circles.  A majority of the lower velocity galaxies (9/10) occupy a specific region of the sky along the line that bisects the WAT.  Such velocity data may be an indication of some portion of a cluster merger occurring along the line-of-sight with a radial velocity of $\sim$ 1500 km s$^{-1}$.  The contours are the same as those in Fig. 2.}
	\end{figure}

    \begin{figure}
      \plotone{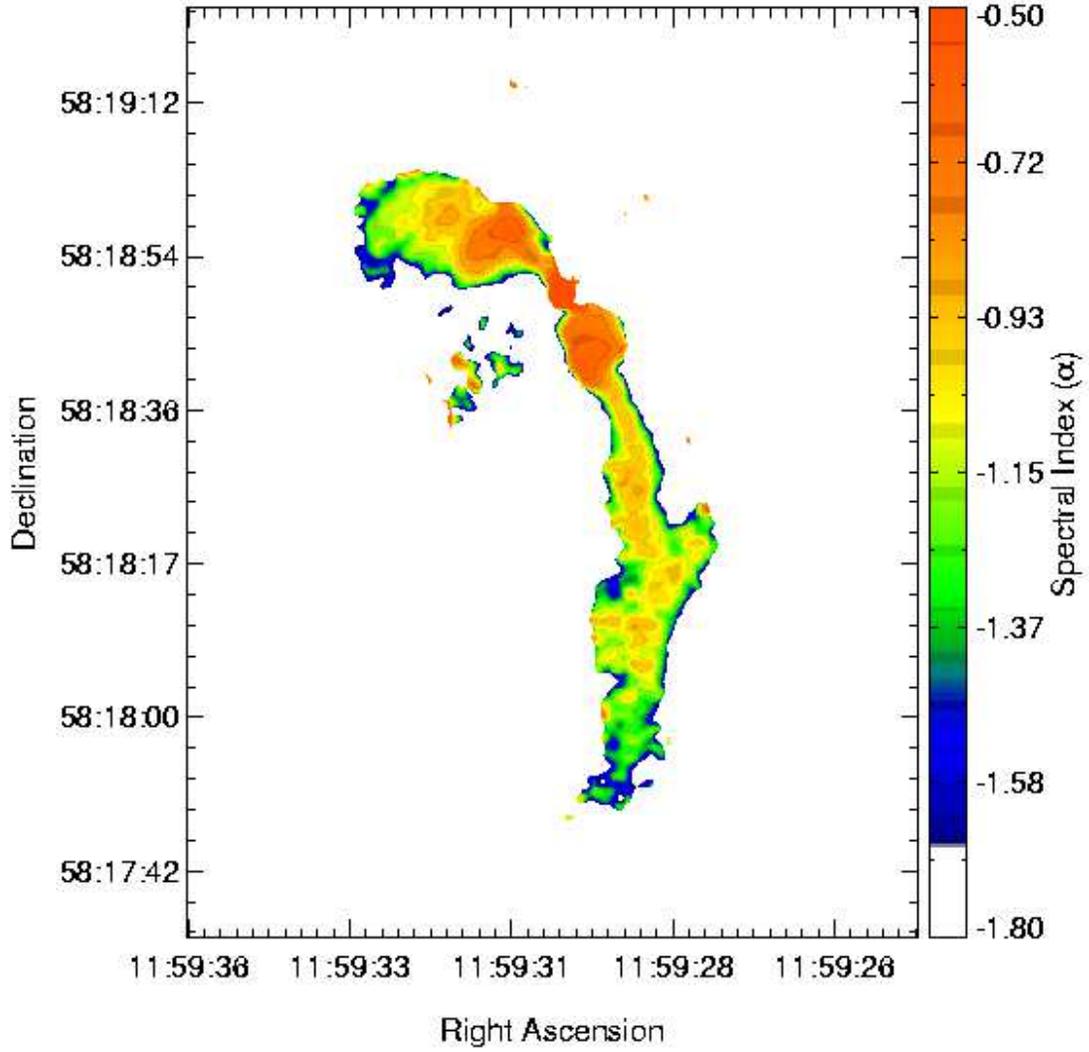}
      \caption{\label{fig:sed} Spectral index map of 1159+583 obtained from 1.4 GHz and 4.8 GHz VLA data.}
	\end{figure}

    \begin{figure}
\epsscale{1}
      \plotone{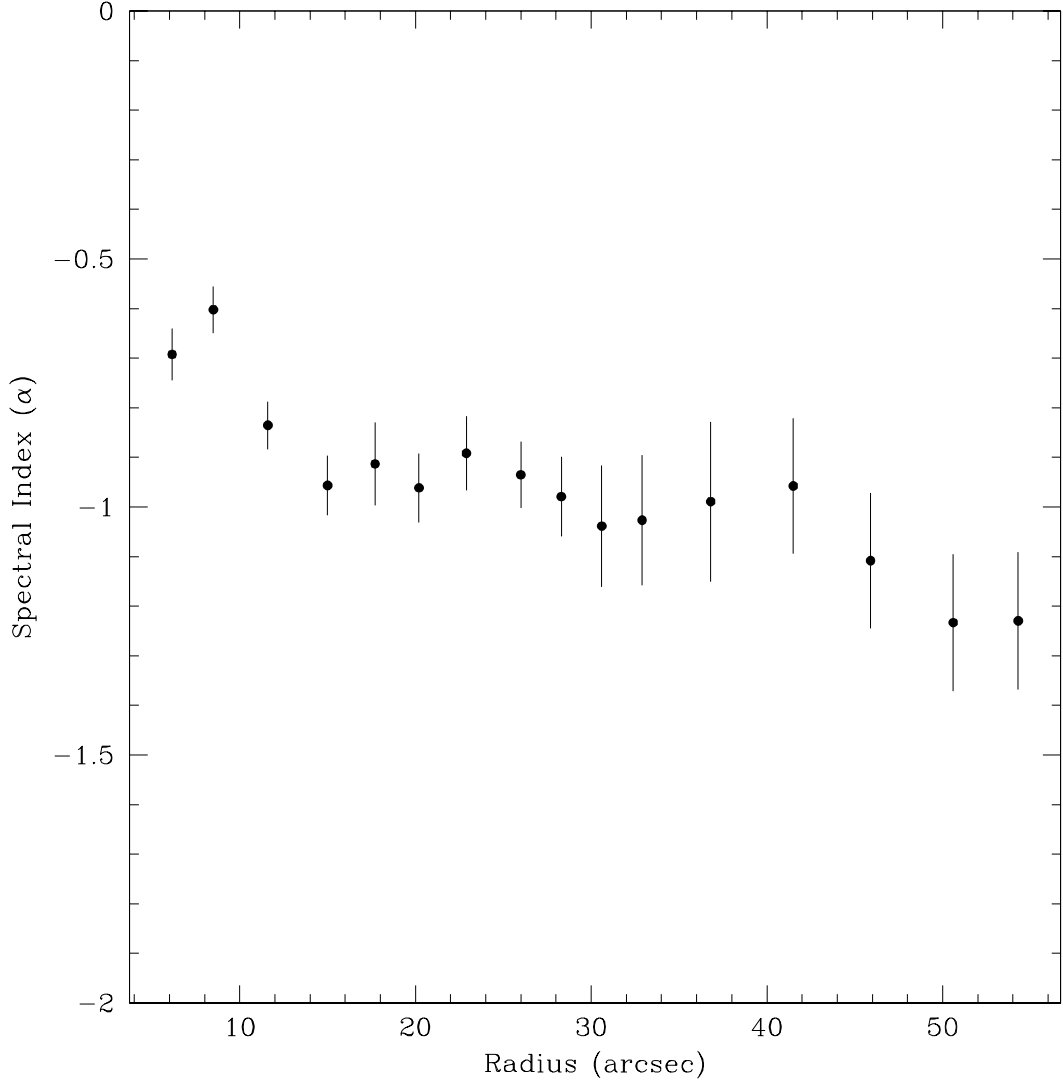}
      \caption{\label{fig:sed} Radial profile of the spectral index of the southern lobe of 1159+583 obtained from 1.4 GHz and 4.8 GHz VLA data.}
	\end{figure}

   \begin{figure}
	\epsscale{0.8}
      \plotone{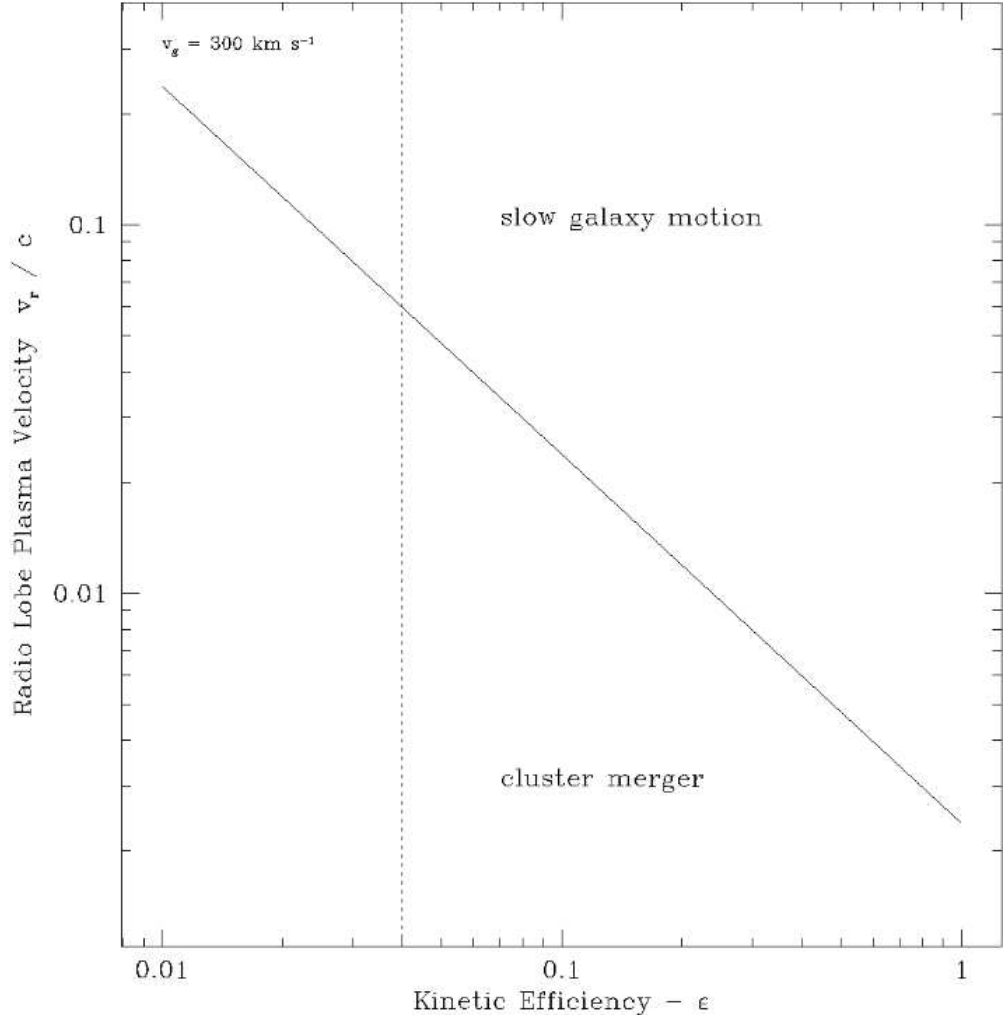}
      \caption{\label{fig:sed} Plasma flow speed down the lobes versus kinetic efficiency.  The diagonal line represents the combinations of plasma flow speed and kinetic efficiency required to bend the lobes into their observed shape given a galaxy velocity of 300 km s$^{-1}$ (the upper limit for systematic galaxy motion about the cluster center).  Combinations that fall below this line require galaxy velocities greater than 300 km s$^{-1}$, while those that lie above allow velocities less than 300 km s$^{-1}$.  The dotted line corresponds to our calculated kinetic efficiency ($\epsilon \sim 0.04$).}
	\end{figure}

\end{document}